\documentclass[twoside,11pt]{article}
\usepackage[dvipsnames]{xcolor}

\usepackage{custom}

\usepackage{algorithm, algorithmicx}
\usepackage{algpseudocode}
\usepackage{subfig}
\usepackage{physics}
\usepackage{subcaption}
\usepackage{float}
\usepackage{graphicx}
\usepackage{graphicx} 
\usepackage{adjustbox}

\usepackage{amsfonts}
\usepackage{amsmath}
\usepackage{amssymb}
\usepackage{mathtools}

\usepackage{pstricks, pst-plot}	
\usepackage[figuresright]{rotating}

\usepackage[english]{babel}
\usepackage{blindtext}
\usepackage{braket}

\DeclareMathOperator*{\argmax}{argmax}
\DeclareMathOperator*{\argmin}{argmin}

\title{Towards Efficient Quantum Computation of Molecular Ground State Energies using Bayesian Optimization with Priors over Surface Topology}

\author[1]{Farshud Sorourifar}
\author[4,5]{Mohamed Taha Rouabah}
\author[4,5]{Nacer Eddine Belaloui}
\author[4,6]{Mohamed Messaoud Louamri}
\author[7]{Diana Chamaki}
\author[2]{Erik J. Gustafson}
\author[3]{Norm M. Tubman}
\author[1]{Joel A. Paulson}
\author[2,3,8]{David~E.~Bernal~Neira }

\affil[1]{\RaggedRight Department of Chemical and Biomolecular Engineering, The Ohio State University, Columbus, OH, USA}
\affil[2]{USRA Research Institute for Advanced Computer Science, Mountain View, CA, USA}
\affil[3]{NASA Ames Research Center, Moffett Field, CA, USA}
\affil[4]{Constantine Quantum Technologies, University Constantine 1 Fr\'eres Mentouri, Constantine, Algeria}
\affil[5]{Laboratoire de Physique Math\'ematique et Subatomique, University Constantine 1 Fr\'eres Mentouri, Constantine, Algeria}
\affil[6]{Theoretical Physics Laboratory, University of Science and Technology
Houari Boumediene, Algiers, Algeria}
\affil[7]{Department of Chemistry, Columbia University, New York, NY, USA}
\affil[8]{Davidson School of Chemical Engineering, Purdue University, West Lafayette, IN, USA} 

\email{dbernaln@purdue.edu}

\begin{document}
\maketitle
\thispagestyle{empty}  

\begin{abstract}
Variational Quantum Eigensolvers (VQEs) represent a promising approach to computing molecular ground states and energies on modern quantum computers.
These approaches use a classical computer to optimize the parameters of a trial wave function, while the quantum computer simulates the energy by preparing and measuring a set of bitstring observations, referred to as shots, over which an expected value is computed.
Although more shots improve the accuracy of the expected ground state, it also increases the simulation cost. 
Hence, we propose modifications to the standard Bayesian optimization algorithm to leverage few-shot circuit observations to solve VQEs with fewer quantum resources.
We demonstrate the effectiveness of our proposed approach, Bayesian optimization with priors on surface topology (BOPT), by comparing optimizers for molecular systems and demonstrate how current quantum hardware can aid in finding ground state energies. 
\end{abstract}

\Keywords{Variational Quantum Eigensolver, Bayesian Optimization, Gaussian Processes}

\section{Introduction} 
\label{sec:intro}
Solving electronic structure problems for molecular systems is a cornerstone of modern chemical engineering research. These problems involve solving Hamiltonian eigenvalue systems, which are used to define the ground-state energy, the lowest possible configuration of a molecular system, which directly influences the behavior of the system~\cite{ciapprox-2,ciapprox-book}. These energies are used to predict molecular properties \cite{Faber17}, understand chemical reactions \cite{Li2024, Dietschreit}, control molecular systems \cite{Nodozi,Findeisen}, as well as design pharmaceuticals \cite{BOONE2016221}, polymers \cite{Arora,Chen2024},  materials \cite{Bui2024, Bombarelli2016, BEFORT20221249,ALEXEEV2024666,Bassman2021}, and catalysts \cite{Kwon,Goldsmith,NIKBIN201335}. Consequently, these calculations affect the design and optimization of industrial processes, with applications spanning healthcare, energy, and the environment \cite{Adjiman21}.

Calculating the ground state involves solving the Schr\"{o}dinger equation, which can only be solved exactly for a few selected cases. The interactions between increasingly many electrons result in an exponential computational complexity, making exact solutions practically intractable for large systems \cite{Derezinski, Guzik2005}. Although approximate methods are used to compute the ground state~\cite{eriksen2020ground, Kim_2018,ciapprox-1}, the scaling issue with many-body calculations persists, requiring further approximations for larger systems \cite{Bogojeski2020}. Furthermore, these approximations are known to erode the quality of the solutions to systems with strong electron-electron interactions \cite{Shi23}.   

Solving many-body systems is a problem class where quantum computers are expected to excel, making them a promising tool for efficiently computing exact ground state energies \cite{TILLY20221, Fauseweh2024,gustafson2024surrogate, Grimsley_2023, Anschuetz_2022}, which also includes hardware demonstrations~\cite{Kandala2017,gustafson2024surrogate,PRXQuantum.3.020323,PhysRevResearch.5.033071,darbha2024false,darbha2024long,2024arXiv24beyond,arute2020observ,ollitrault2024estimation}. An algorithm designed explicitly for this task is known as the variational quantum eigensolver (VQE), which has been used to great effect in many problems \cite{anastasiou2022tetrisadaptvqe, gustafson2024surrogate, 2016NJPh...18b3023M,bittel2022fast, PhysRevA.98.032309, Harrow2021,jones2020efficient, Berg2022, zhew2020, PhysRevA.107.032415,hirsbrunner2024diag,leimkuhler2024,ollitrault2024enhancing,2024arXivruslan,Fin_gar_2024,Cheng_2024}. This approach leverages the quantum computers' capacity for efficiently simulating many-body systems, which can be used to determine the ground state energy, $E_0$, from the variational principle of quantum mechanics, 

\begin{equation*}\label{eq:variational_principle}
    E_0 \leq  \frac{\bra{ \Psi (\theta)} \mathcal{H}  \ket{\Psi (\theta)}}{\bra{\Psi (\theta)}\ket{\Psi (\theta)}}.
\end{equation*}

Here, $\Psi(\theta)$ is a trial wave function of the system parameterized by $\theta$, and $\mathcal{H}$ is the Hamiltonian acting on the system. Note that in practice, the trial wave functions may be normalized so that  $\bra{\Psi (\theta)}\ket{\Psi (\theta)}=1$. The trial wave function represents the state of the electronic configuration, whereas the Hamiltonian defines the system's energy. The VQE determines $E_0$ by formulating equation \eqref{eq:variational_principle} as an optimization problem, 
\begin{equation*}\label{eq:vqe_objective}
    \min_{\theta \in \mathbb{R}^d} y(\theta) = \frac{1}{S}\sum^S   \frac{\langle \Psi (\theta)| \mathcal{H}|   \Psi (\theta)\rangle}{\langle \Psi (\theta)|\Psi (\theta)\rangle}
\end{equation*}
which is solved on a non-quantum or \emph{classical} computer using optimization by simulating $y(\theta)$ on the quantum computer for selected $\theta$. The sample-based expectation value is calculated over $S$ repeated observations of the final quantum state, referred to as shots. At the same time, this theoretically allows for arbitrary precision, although a greater number of measurements naturally incurs additional costs.

Recently, Bayesian optimization (BayesOpt) has been investigated as a means of solving Equation \eqref{eq:vqe_objective} using VQE and has shown competitive performance against other commonly used optimization routines \cite{otterbach2017unsupervised,tibaldi23,Ciavarella22,iannelli2021noisy}. BayesOpt is a general class of model-based derivative-free optimization (DFO) algorithms. Model-based DFO methods use a surrogate model to determine a sampling strategy, in contrast to model-free strategies (e.g., genetic algorithms), which are less sample-efficient \cite{Eriksson19}. What differentiates BayesOpt from other DFO methods is the use of a statistical surrogate to model a mean-based prediction and the epistemic uncertainty (uncertainty caused by limited data) to determine the most efficient sampling strategy. Gaussian processes (GPs) are frequently used due to their excellent capacity for uncertainty quantification, assuming the kernel function, which defines the GP model, has been appropriately selected \cite{Rasmussen2006, nei, jones1998efficient,balandat2020botorch,frazier18}. The mean and uncertainty functions derived from the surrogate model are used to trade off between exploring the parameter space (i.e., sampling in regions of greater uncertainty) and exploiting the most promising regions (e.g., sampling near the best-observed parameters).
This exploration and exploitation trade-off allows the global optimum to be found with limited evaluations.

BayesOpt for VQE problems has been the subject of growing research interest, with most work focused on using standard implementations with limited modifications \cite{otterbach2017unsupervised, tibaldi23, Ciavarella22, iannelli2021noisy}, or use BayesOpt as a component of a larger optimization framework \cite{Muller22, Tamiya2022}. 
Several trust-region-based BO methods have also been investigated \cite{Shaffer2023, Cheng2024}, which provide convergence stability and allow for parallel sampling. 
In \cite{Self2021}, the authors propose a parallel BayesOpt strategy for solving related problems, such as solving the VQE for an ensemble of molecular geometries, by sharing sample data across the different problems, and \cite{rohrs2024bayesian} builds on this work by investigating different information sharing schema. 
The only work of which we are aware to propose modifications to the GP model specialized for VQE is in \cite{Nicoli2023} where the authors develop a novel periodic kernel to construct the GP model, in addition to a novel acquisition function for their trust-region based BaysOpt framework, inspired by the Nakanishi-Fuji-Todo method \cite{Nakanishi2020}.

In this work, we build on our proposed BayesOpt framework \cite{escape}, which constructs a prior on surface topology from noisy low-shot observations and then optimizes for high-shot observations (i.e., the number of shots required to achieve a prescribed accuracy) using a residual GP model. The prior on surface topology uses a data-driven surrogate model as an initial belief of $y(\theta)$ in lieu of the traditional zero-mean prior. Additionally, we find that encoding information about the periodicity of the rotation gate parameters translates into faster convergence. We modify the prior on surface topology to accommodate larger datasets and consider the practical usage of quantum computers. 
Specifically, we investigate the algorithm's performance for three noise models constructed from existing IBM quantum computers (computational results) and on physical quantum hardware (experimental results).
While current quantum hardware tends to provide biased estimates of the ground state, BayesOpt can still find optimal circuit parameters, suggesting quantum computers may be useful for determining optimal circuit parameters, i.e., $\theta^\star \in \argmin_{\theta \in \mathbb{R}^d} y(\theta)$ in Equation \eqref{eq:vqe_objective}, and computing the energy classically. The rest of the paper is organized as follows: we first provide a detailed background on quantum computing, variational quantum eigensolvers, and Bayesian optimization in Section \ref{sec:prelims}. We formally introduce the problem in Section \ref{sec:problem-formulation} and present our methods in Section \ref{sec:methods}. In Section \ref{sec:casestudies}, we present the VQE convergence results for our method and several benchmarking methods for two molecular systems, with a particular focus on how the algorithms perform on high-fidelity noise models and when using noisy physical quantum computers. In addition, that section discusses strategies for efficiently using currently available quantum hardware to compute molecular ground-state energies relative to those of purely classical methods. Lastly, we provide concluding remarks and directions for future work in Section \ref{sec:conclusion}.

\begin{figure}
    \centering
    \includegraphics[width=\textwidth]{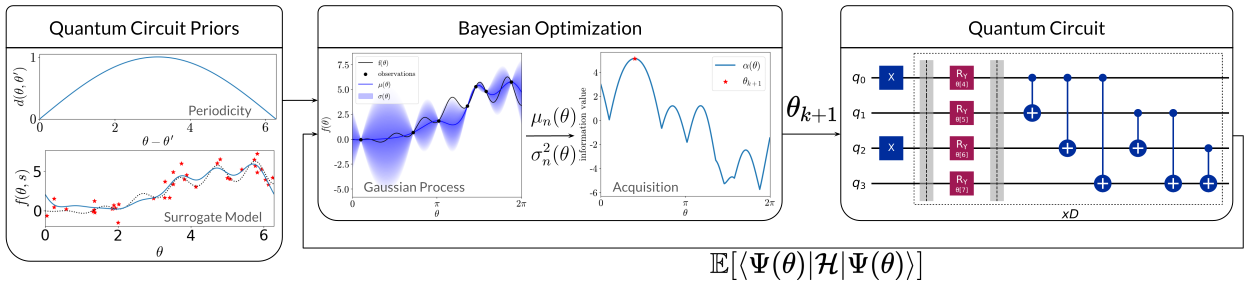}
    \caption{{\bf{B}}ayesian {\bf{O}}ptimization with {\bf{P}}riors on {\bf{T}}opology (BOPT) Summary. Priors on the quantum circuit parameter's $2\pi$ periodicity and surface topology, in the form of a periodic kernel and surrogate models, are used to initialize the Bayesian Optimizations loop. At each iteration, the Gaussian process is conditioned on the available circuit observations, which can be used to construct the acquisition function. Optimizing the acquisition provides the subsequent circuit parameters, which are used to query the circuit. With the new observation, the loop is repeated until the sample budget is exhausted. }
    \label{fig:BOPT_summarylabel}
\end{figure}


\section{Background and Preliminaries}\label{sec:prelims}
\subsection{Quantum computing}\label{sec:prior_qc}

Quantum information processing is a strategy to process information in accordance with the principles of quantum mechanics. One area of quantum information processing, quantum simulation, is used to solve problems formulated as quantum systems. Many problem classes can be modified to fit these criteria, such as factoring large numbers (for cryptographic applications), simulating quantum systems, and solving optimization problems \cite{Bernal22}. Although it is important to note that quantum simulation is not necessarily superior to classical methods, there are specific cases where it is expected to excel, such as many-body problems \cite{Fauseweh2024}.

Quantum computers, or quantum computing units (QPU), exploit quantum mechanical phenomena to process information. This information processing is performed physically using systems that exhibit multi/two-level quantum behavior, such as superconducting qubits \cite{Castelvecchi2017,Linke} present in IBM quantum devices. These two-qubit states are known as \emph{qubits}. For quantum computers, the computations are interpreted as simulations. These simulations are represented by evolving a unit vector in the Hilbert space, which represents the quantum state, based on the rules of quantum mechanics. Although computationally expensive, these simulations can also be performed on classical computers, where quantum effects are implemented by exponentially growing states with the number of qubits. Conversely, leveraging quantum mechanical phenomena directly in a quantum computer has the potential to scale much more favorably since the computation is replaced by a controlled simulation of the physical system in the computer.  

Specifically, the quantum phenomena that are advantageous for quantum simulations and improved information processing are superposition and entanglement. Unlike classical bits, which only exist in one of two states (0 or 1), qubits, the fundamental unit of quantum information, can be in a coherent superposition of both states simultaneously. The superposition of states allows more information to be encoded in the base computing unit. Entanglement is a phenomenon where the state of one qubit is intrinsically correlated with the state of another qubit so that states cannot be factored into products of individual states.
In quantum computing, this allows information that grows exponentially with the number of qubits to be processed seamlessly. In addition, many-body systems often exhibit entanglement among their constituent particles, making it a useful phenomenon for quantum simulation. For a detailed discussion on the topic, we point the interested reader to \cite{nielsen2010quantum}.

A well-defined initial state for each orbital in a molecular quantum system is prepared to simulate it.
These states are propagated through a quantum circuit and are manipulated by quantum gates. The gates perform unitary transformations, such as parameterized rotations and controlled gates, to modify the quantum state. The set of gates or \emph{circuit} is designed to represent the trial wave function, where the process of transforming the initial state is known as state preparation. After the circuit execution, when the final state of the circuit is obtained, it is \emph{measured}. In this process, the coherent state is projected or \emph{collapsed} onto a classical state represented by a string of bits, e.g., 00110101.
Each single execution is known as a \emph{shot}, which is then mapped to an observable via an operator, which in this case is the energy computed through the Hamiltonian. Lastly, a sample-averaged expectation is taken over the collection of observed values. On quantum hardware, running a shot is fast, and problems with 27 qubits using thousands of shots can be simulated in seconds \cite{Dahlhauser}.  

Although quantum computing is a promising and exciting technology for simulating quantum systems, we emphasize that current computers are noisy intermediate-scale quantum (NISQ) devices with open technical challenges \cite{Preskill2018quantumcomputingin}. The largest current quantum computers still have too few and noisy qubits ($\sim 500$) 
\footnote{If not for current levels of noise, $\sim$ 100 qubits would provide sufficient computing resources for the utility of quantum computing \cite{kim2023evidence}. However, NISQ error correction schemes often call for additional qubits \cite{shor1995a,Kandala2019}.}
, both limiting the size of molecular systems that can be simulated and requiring the use of hybrid quantum-classical algorithms, like VQE, that offload some of the computational burdens onto classical devices.
The quantum states of qubits are also fragile and prone to decoherence because of environmental disturbances, such as thermal fluctuations. As a result, the quantum states in NISQ devices can only be maintained for short periods before decohering, limiting the complexity of circuits that can be implemented and of problems that can be solved. Additionally, noise is introduced due to hardware limitations, such as inaccuracies when implementing gate protocols. While fully characterizing the quantum state is subject to the inherent uncertainty of quantum mechanics, this can be addressed by using a large number of shots to observe the circuit, which incurs additional costs. Thus, the practical value of modern quantum computers depends on creative solutions for navigating noisy observations and minimizing the computational load assigned to the quantum computer.

\textbf{Remark:}
While there is a myriad of quantum computing technologies that are continuously being improved, we focus on a topical treatment of the subject. The goal of this section is to provide high-level connections between components in common hardware architectures and explain how they relate to the mathematical concept of quantum simulation. While we strive to remain technology agnostic, for brevity, we resort to gate-based models of quantum computers to describe the process of simulating a trial wave function. We note that the gate-based QPU architectures are one of the most widely studied quantum computers and constitute the computational architecture used in our hardware simulations.

\subsection{Variational Quantum Eigensolvers}\label{sec:prior_vqe}
The VQE algorithm uses a hybrid quantum-classical approach to determine the ground state of molecular systems by optimizing the parameters of a trial wave function $\Psi(\theta)$, which is simulated in the quantum device (i.e., either a quantum computer or classical quantum simulator). 
First, a reference state $\Psi_{\text{ref}}$ is selected to initialize the trial wave function, such as the Hartree-Fock reference state.
The principal challenge in VQE is known as state preparation, which is the process of designing a trial wave function to minimize the expected energy.
This step can be separated into two parts: the selection of a circuit structure known as \emph{ansatz} that represents the trial wave function and the selection of parameters of the ansatz to
drive the system from the reference state to the trial state exhibiting the lowest energy configuration.     

Although the ansatz selection is still an open question, we present two that are frequently encountered: those from the class of hardware efficient ans\"atze (HEA) \cite{Kandala2017, Wu2021, xiao2023physicsconstrained, peruzzo2014variational, anastasiou2023really}, and the unitary coupled cluster with singles and doubles (UCCSD) ansatz \cite{romero2018strategies, Cao2019, Anand_2022}.
In addition, the circuit depth $D$, which is the number of repeating layers, is often treated as part of the ansatz.
Deeper circuits result in increased representational capabilities, ensuring that $\Psi(\theta)$ can model the ground state but also increase the circuit running time and the likelihood of decoherence.

UCCSD is a chemistry-inspired ansatz rooted in classical coupled cluster theory, a well-established method for computing electronic structures of molecules.
The method is designed to capture the most important electronic correlations in the molecular system by applying single- and double-excitation operators to the reference state.
Formally, the USSCD ansatz is defined as, 
\begin{equation}
    \ket{\Psi({\theta})} =  \big[ \Pi_{i=1}^D e^{\hat{T}({\theta}) - \hat{T}({\theta})^{\dag}} \big] \ket{\Psi_{ref}},
\end{equation}
where $\theta$ represents the variational parameters that need to be optimized, and $\hat{T}({\theta}) = \hat{T}_1({\theta}) + \hat{T}_2(\vec{\theta})$ are the aggregated single and double excitation operators.

\begin{figure}[h]
    \centering
    \includegraphics[width=.55\textwidth]{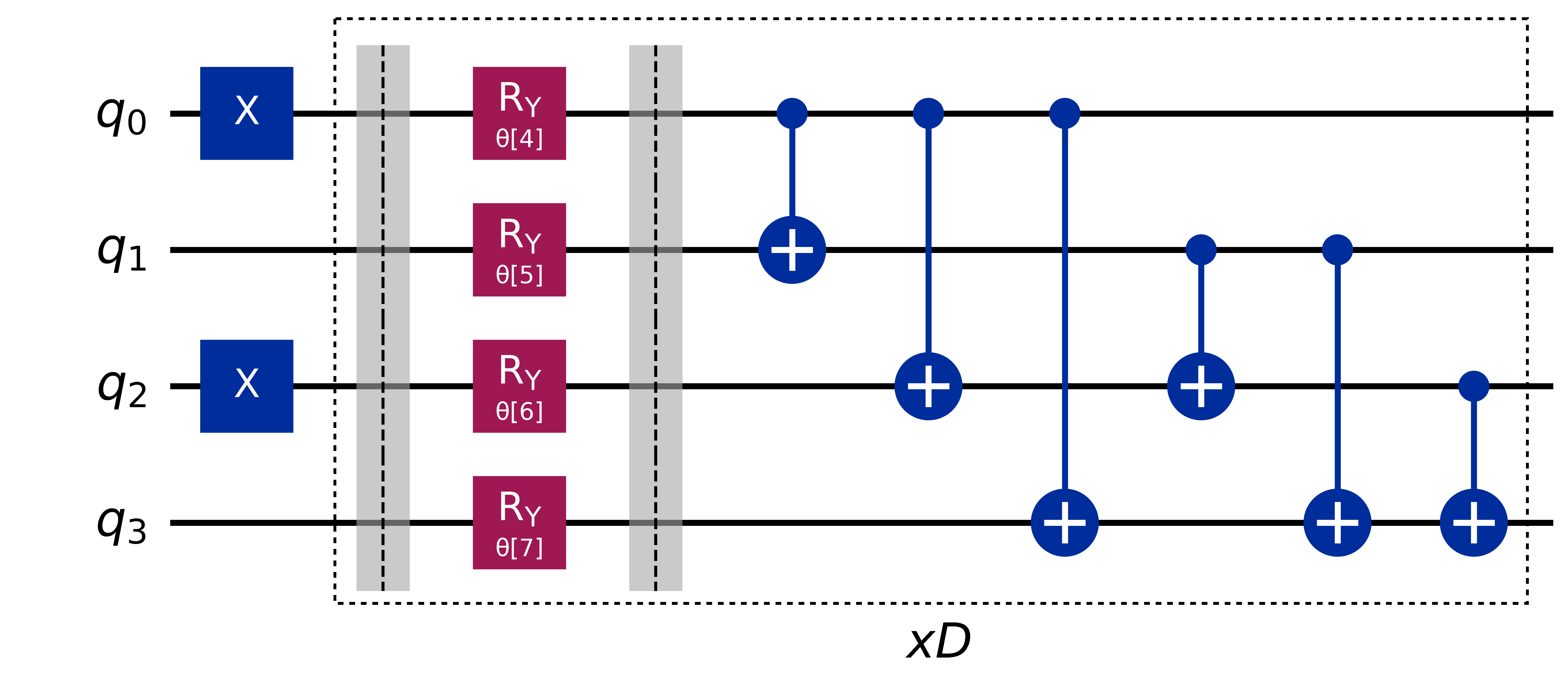}
    \hfill
    \includegraphics[width=.35\textwidth]{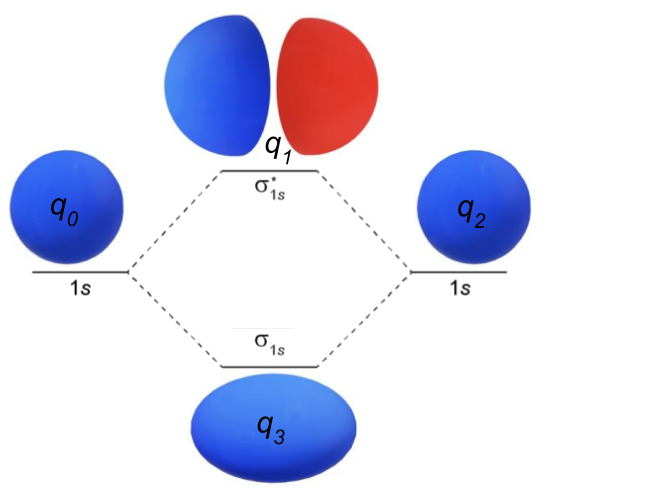}
    \caption{(Left) Hartree-Fock initialization (shown in blue boxes with ``X") HEA for $H_2$ (4 qubit system) using $D$ layers composed of $R_y$ single qubit gates and CNOT entanglement gates acting on all possible qubit pairs. The dashed box denotes the layer. (Right) $H_2$ molecular orbital diagram showing the orbital qubit assignments.    }
    \label{fig:HEA_circuit}
\end{figure}

The HEA, as its name suggests, is a more flexible class of ans\"atze designed to use fewer gates and use NISQ-era QPUs more efficiently.
As a result, they are not limited to chemical systems.
While these features sound appealing, it is unclear whether tailored ans\"atze can potentially provide better utility with a larger number of gates, e.g., by requiring fewer layers.
A generic HEA can be defined as  
\begin{equation}
    \ket{\Psi(\vec{\theta})} = \big[ \Pi_{i=1}^D \hat{\mathrm{U}}_r(\theta_i) \times \hat{\mathrm{U}}_{\text{ent} }\big]\ket{\Psi_{\rm ref}},
\end{equation}
where $\hat{\mathrm{U}}_{\text{ent}}$ are the entangling gates, and $\hat{\mathrm{U}}_r(\theta_i)$ are the single qubit gates.
Here, $\theta_i\in\mathbb{R}^{n_g}$ is the parameter vector for each layer, where $n_g$ is the number of gates in each layer, such that $d = n_gD$.
We present a schematic overview of a quantum circuit using the HEA in Figure \ref{fig:HEA_circuit}.
For more details on VQE algorithms and best practices, see \cite{TILLY20221}.

\subsection{Bayesian Optimization}\label{sec:prior_bo}

As mentioned previously, BayesOpt is a model-based optimization algorithm geared toward expensive black-box functions.
It consists of two key components: a statistical surrogate model and an acquisition function, the latter of which defines how to use the model to identify subsequent sample points.
The algorithm proceeds sequentially.
The GP is fit to available data, and the predicted mean and variance are used to construct an acquisition function.
The acquisition function is maximized to determine the following sample location.
The black-box function is evaluated at the selected sample point, which is then added to the data set, and the process is repeated until a specified sample budget $B$ is exhausted.
In practice, a fraction of the initial budget $B_{\text{init}}$ is typically allocated to pseudo-random sampling to generate an initial data set when little to no prior data are available.

We begin by introducing the mathematical underpinning of GPs and their use in BayesOpt; interested readers are referred to \cite{Rasmussen2006} for further details. GPs generalize the notion of a multivariate Gaussian to function space, representing an uncountable set of functions that fit the data. Specifically, it constructs a posterior probability distribution over function space that can be conditioned on the observed data, in accordance with Bayes's rule,
\begin{equation}
    P\left(f(\theta)|\mathcal{D}^\star_n\right) \propto P\left(\mathcal{D}^\star_n|f(\theta)\right)P\left(f(\theta)\right).
\end{equation}
Here, $P(f(\theta))$ denotes the prior probability of observing $f$ for a given $\theta$, and $P(\mathcal{D}^\star_n|f(\theta))$ denotes the conditional probability of observing the data $\mathcal{D}^\star_n = \{\left(\theta_i,f(\theta_i)\right)\}_{i=0}^{n}$ (also known as the likelihood function). The key assumption in a GP, denoted by $GP(\mu(\theta), k(\theta,\theta'))$, is that the function values at any subset of inputs are jointly Gaussian such that they are fully defined by their mean function $\mu: \mathbb{R}^d \to \mathbb{R}$ and covariance or \emph{kernel} function $k: \mathbb{R}^d \times \mathbb{R}^d \to \mathbb{R}_{\geq 0}$.

In general, $f(\theta)$ cannot be directly observed, and instead, the measurement process introduces a source of noise that can be modeled as $y = f(\theta) +\epsilon$, where $\epsilon\sim\mathcal{N}(0,\sigma^{2})$. Given a noisy dataset $\mathcal{D}_n=\{(\theta_i,y_i)\}_{i=0}^{n}$, the posterior probability of $f(\theta)$ can be modeled jointly with the noisy observations $y_{1: n}$ as
\begin{equation}
\left[\begin{array}{l}
y_{1: n} \\
f(\theta)
\end{array}\right] \sim \mathcal{N}\left(\left[\begin{array}{c}
{\mu_0}(\theta_{1:n}) \\
\mu_0(\theta)
\end{array}\right],\left[\begin{array}{cc}
\mathbf{K}_n+\sigma^2 I_n & \mathbf{k}_n(\theta) \\
\mathbf{k}_n(\theta)^{\top} & k_0(\theta, \theta)
\end{array}\right]\right).
\end{equation}

Here, $\mathbf{k}_n(\theta) \in \mathbb{R}^n$ is a vector of covariance terms between the observed inputs $\theta_{1:n}$ and the test point $\theta$, i.e., $[\mathbf{k}_n(\theta)]_i = k(\theta_i,\theta)$,
and $\mathbf{K}_n \in \mathbb{R}^{n \times n}$ is a symmetric covariance matrix with elements $[\mathbf{K}_n]_{i,j} = k(\theta_i,\theta_j)$ for all $i,j = 1,\ldots, n$.
Due to the properties of joint Gaussian random variables, the posterior distribution $p(f(\theta) |\mathcal{D}_n, \theta)$ given all available observations, is normally distributed with the following posterior mean and variance as a function of the test point
\begin{subequations} \label{eq:posterior-mean-variance}
\begin{align}
\mu_{n}(\theta) &= \mu_0(\theta) + \mathbf{k}_n(\theta)^\top (\mathbf{K}_n + \sigma^2 I_n ) ( y_{1:n} - {\mu}_0(\theta_{1:n}) ),\label{eq:gp_mean} \\
\sigma^2_{n}(\theta) &= k_0(\theta,\theta) - \mathbf{k}_n(\theta)^\top (\mathbf{K}_n + \sigma^2 I_n)^{-1} \mathbf{k}_n(\theta).\label{eq:gp_var}
\end{align}
\end{subequations}
Note that here, $\mu_0$ and $k_0$ refer to the prior mean and kernel functions, respectively.
The matrix inverse $(\mathbf{K}_n + \sigma^2 I_n)^{-1}$ in equation \eqref{eq:posterior-mean-variance} is often considered a limiting factor for GP models - the computational complexity scales cubically with the number of data points, making inference computationally challenging for large data sets.

A zero-mean prior ($\mu_0(\theta) = 0$) is generally a reasonable assumption when normalizing the data to be zero-centered.
The kernel function is typically selected based on a prior belief about the degree of smoothness of the underlying function.
A common choice of the kernel class is that from the Mat\'ern class,
\begin{equation}
k_\text{Mat\'ern}\left(\theta, \theta^{\prime}\right)=\sigma_k^2 \frac{2^{1-\nu}}{\Gamma(\nu)}\left(\sqrt{2 \nu} {\sum_{i=1}^d \rho_i\left(\theta_i-\theta_i^{\prime}\right)}\right)^\nu K_\nu\left(\sqrt{2 \nu} {\sum_{i=1}^d \rho_i\left(\theta_i-\theta_i^{\prime}\right)}\right),
\end{equation} 
where $\Gamma$ is the gamma function and $K_\nu$ is the modified Bessel function of the second kind.
The parameters $\rho_i$ for all $i = 1,\ldots, d$ are inverse squared length scale parameters, which models the response sensitivity for each dimension of $\theta \in \mathbb{R}^d$, and $\sigma_k^2$ is a scale parameter meant to capture expected variation in $f$.
The parameter $\nu$ controls the smoothness of the GP model, with an integer value $\nu$ resulting in a model that is $\nu-1$ time differentiable.
In the limit $\nu\rightarrow\infty$, the well-known squared exponential (or radial basis) function is recovered.  

Another much less commonly used kernel is the periodic kernel.
Considering that the parameters used in quantum circuits are periodic, and this kernel is a natural choice.
The periodic kernel is defined as 
\begin{equation}
k_{\text {Periodic}}\left(\mathbf{\theta}, \mathbf{\theta}^{\prime}\right)=\sigma_k^2\exp \left(-2 \sum_i {\sin ^2\left(\frac{\pi}{p}\left(\theta_i-\theta_i^{\prime}\right)\right)\rho_i}\right).
\end{equation}
While $p$ is typically an unknown parameter, in VQE, setting $p=2\pi$ for $\theta_i\in[0,2\pi]$ can be used to enforce periodicity.
Furthermore, we note that this kernel can be considered another topological prior, as it represents the relationship between the cost function and the circuit parameters, which is a sine curve with period $2\pi$ \cite{Nakanishi2020}. 

Part of selecting the kernel also requires choosing the hyperparameters $\gamma=\left\{\rho_1, \ldots, \rho_D, \sigma_k^2 \right\}$.
This selection is made after each new observation by minimizing the negative marginal log-likelihood, 
\begin{equation}\label{eq:marginal-log-likelihood}
\gamma^\star = \argmin_\gamma \underbrace{\frac{1}{2}(\mathbf{y}_n - \boldsymbol{\mu}_n)^{\top} \Sigma_{\mathbf{y}_n}^{-1}(\mathbf{y}_n-\boldsymbol{\mu}_n)}_{\text {Data fit term }}+\underbrace{\frac{1}{2} \log \left(\left|\Sigma_{\mathbf{y}_n}\right|\right)}_{\text {Complexity penalty }}+\frac{n_y}{2} \log (2 \pi). 
\end{equation}
where $\mathbf{y}_n = y_{1:n}$, $\boldsymbol{\mu}_n = \mu(\theta_{1:n})$, and  $\Sigma_{\mathbf{y}_n}= \mathbf{K}_n+\hat{\sigma}_f^2 I_n$. 
The negative marginal log-likelihood is a loss function that provides some resilience to overfitting.
The data-fit term encourages the model to match the available data, while the complexity term provides a regularization effect that encourages a more simple model. 

For a given dataset, we can use equation \eqref{eq:marginal-log-likelihood} to fit a GP to a dataset, which provides a mean and covariance function defined by equations \eqref{eq:posterior-mean-variance}.
Now, we turn to the acquisition function, which is used to select subsequent samples in search of optimal circuit parameters. 
One clear example of how \eqref{eq:posterior-mean-variance} can be used to select $\theta_{n+1}$ to balance competing objectives (exploration versus exploitation) is the lower confidence bound (LCB) acquisition function 
\begin{equation}
\alpha_{\text{LCB}, n}(\theta) = \underbrace{\mu_n(\theta)}_{\text{Exploitation}} - \underbrace{\sqrt{\beta}\sigma_n(\theta)}_{\text{Exploration}}, 
\end{equation} 
which determines the new sample point by minimizing $\alpha_{\text{LCB}, n}(\theta)$ over some compact candidate set $\Theta \subseteq \mathbb{R}^d$
\begin{equation}
    \theta_{n+1} = \argmin_{\theta \in \Theta} \alpha_{\text{LCB}, n}(\theta),
\end{equation}
which can be solved with standard gradient-based optimization.
Here, the mean function promotes the exploitation of regions that are known to be good, while the covariance function, scaled by the confidence interval parameter $\sqrt{\beta}$, promotes exploitation.
Together, the acquisition function assigns optimistic values to each candidate point. 

Furthermore, we introduce the more commonly used expected improvement acquisition function, which measures the amount by which the current objective is expected to improve over the current best or \emph{incumbent} solution, $\eta$.
Formally, the expected improvement criterion for minimizing an objective function is defined as
\begin{equation}\label{eq:ei_criterion}
\alpha_{\text{EI},n}(\theta)=\mathbb{E}_n\big[\max \left(\eta - f(\theta), 0\right)\big].
\end{equation}
The EI acquisition function is commonly used since an analytic solution can be derived for GP models \cite{jones1998efficient} 
\begin{align} \label{eq:expected-improvement}
\alpha_{\text{EI}, n}(\theta) = \sigma_n(\theta) \left( z_n(\theta) \Phi(z_n(\theta) \right) + \phi\left(z_n(\theta)\right),
\end{align}
where $z_n(\theta) = (\eta - \mu_n(\theta)) / \sigma_n(\theta)$, $\Phi$ is the standard Gaussian cumulative density, and $\phi$ is the probability density. 
While $\eta$ is often set to the best-observed value $\eta = \min_{i \in \{ 1,\ldots,n \}} y_i$, it has been discussed in \cite{wang2014theoretical} that this choice causes issues in a noisy setting.
Since noise is an important characteristic for the objective functions considered in this work, we instead use the best posterior mean value as our incumbent, i.e., $\eta = \min_{\theta \in \Theta} \mu_n(\theta)$.

\section{Problem Formulation}
\label{sec:problem-formulation}
Before introducing the proposed approach, we formally present the VQE problem. 
First, let $f(\theta)$ represent a theoretical circuit evaluated at a given vector of parameters
$\theta$.
The circuit $f(\theta)$ is theoretical in the sense that it represents a noise-free simulation with an exact expectation when observing the state (i.e., in the infinite shot limit). 
We assume that querying the circuit produces observations that are corrupted by the underlying generative process 
\begin{align}\label{eq:circuit_model}
    y_\theta = f(\theta) + \epsilon_{S} + \epsilon_{\text{QPU}}(\theta),
\end{align}
where $\epsilon_{S} \sim \mathcal{N} \left(0,(\sigma_s/\sqrt{S})^2\right)$ is an independent and identically distributed Gaussian noise term, with a standard deviation proportional to $1/\sqrt{S}$, where $S$ is the number of shots chosen to evaluate the circuit (i.e., the standard error in the mean estimate), and the term $\epsilon_{\text{QPU}}, \sim \mathcal{N}\left(\mu_{\text{QPU}}(\theta),\sigma_{\text{QPU}}^2(\theta)\right)$, represents the noise/errors due to hardware. 
Classical quantum simulators can generally be modeled with $\epsilon_{\text{QPU}}=0$.
In contrast, quantum hardware will have an unknown and hardware-specific structure for $\mu_{\text{QPU}}(\theta)$ and $\sigma_{\text{QPU}}^2(\theta)$.
However, data-driven models exist to simulate QPU performance in classical simulations.
Formally, the VQE we aim to solve is 
 \begin{equation}\label{eq:vqe_objective_actual}
    \theta^\star = \argmin_{\theta \in \Theta}  f(\theta)
\end{equation}
where $\Theta \in [0, 2\pi]^d$ is the parameter space. When the number of shots $S$ is sufficiently large, $f(\theta) + \epsilon_{S} \approx f(\theta)$, hence we assume that $S$ is selected to ensure that classically simulated systems are arbitrarily accurate, i.e., $y_\theta \approx f(\theta)$.
This formulation allows for solving equation \eqref{eq:vqe_objective_actual} using observations $y_\theta$.

Although hardware noise mitigation is an active research area, a true noise-resilient QPU at a reasonable scale is probably years away.
As a result, directly observing simulated energies from a QPU does not have much practical value currently, as the errors are often greater than the accuracy required to determine the ground state of these molecular systems, frequently called \emph{chemical accuracy}.
Solving VQEs with classical simulation is computationally prohibitive for systems exceeding $\sim40$ qubits~\cite{2017arXiv170401127H} ($\sim64$ qubits and more are possible if using approximate simulators \cite{mullinax2023large,2023arXiv231012965K}) due to the state preparation.
However, a single simulation with a provided set of circuit parameters is tractable for small molecular systems.
Hence, a possible strategy for the use of near-term quantum computers is to solve equation \eqref{eq:vqe_objective_actual} on a QPU and validate $\theta^\star$ by evaluating \eqref{eq:variational_principle} with a classical quantum simulation.     
 
The form we present in equation \eqref{eq:circuit_model}, using additive noise components, is a useful model; however, the actual noise would enter the system at a variety of intermediate points in the simulation.
For example, gate errors would take the form of $f\left(\theta+\epsilon_{\text{gates}}(\theta)\right)$, where $\epsilon_{\text{gates}}$ is a model of the QPU specific gate implementation error.
The errors introduced from having too few shots are due to a poor sample-averaged approximation of the expectation taken in equation \eqref{eq:vqe_objective}.
However, this form provides a reasonable approximation of the circuit that facilitates more efficient search strategies under the BayesOpt paradigm.

\section{Methods}
\label{sec:methods}
\subsection{Surface Topology Prior}

The proposed framework constructs a prior on the surface topology by fitting a surrogate model to low-shot observations and deploying BayesOpt to learn a residual model between a low-shot surrogate and high-shot observations $y_\theta$.
This framework, which we refer to as a Bayesian optimization with priors on surface topology (BOPT), is inspired by the reference models used in \cite{lu2021bayesian}.

First, we assume that we can generate noisy low-shot observations, 
\begin{align}\label{eq:circuit_model_low_shot}
    \tilde{y}_\theta = f(\theta) + \epsilon_{s} + \epsilon_{\text{QPU}}(\theta),
\end{align}
where $s < S$ is the shot count used to generate noisy observations. 
The costs of querying $y_\theta$ and $\tilde{y}_\theta$ are defined as $1$ and $s / S$ respectively, i.e., the costs are linearly dependent on the number of shots, consistent with commercial QPU cost structures \cite{aws_price, Azure_price}. 
Let $\mathcal{D}^\text{low-shot}_{\text{init}} =\{\theta_i, \tilde{y}_{\theta_i}\}^m$ be the set of $m$ low-shot observations obtained from spending the initialization budget $B_{\text{init}}$ uniformly over the parameter space.
The prior on surface topology is defined in terms of a general surrogate model $\tilde{\mu}_y(\theta)$ for predicting $f(\theta)$ constructed from $\mathcal{D}^\text{low-shot}_{\text{init}}$.
Our proposed Gaussian process with priors on surface topology (GP-TP) can then be expressed as 
\begin{align} \label{eq:gp-tp-prior}
f(\theta) \sim GP( \tilde{\mu}_y(\theta), k_\text{Periodic}(\theta, \theta') ).    
\end{align}
As opposed to using a constant-mean prior function in \eqref{eq:posterior-mean-variance} (the default assumption), we use low-shot observations to learn a more effective prior that shifts the high-shot observations in a nonlinear way. Note that we can actually interpret this process as training a standard zero-mean GP on residual data $\{\theta_i, y_{\theta_i}-\tilde{\mu}_y(\theta_i) \}_{i=1}^n$. The key innovation in BOPT is thus the use of this new type of prior, which we show is capable of speeding up convergence in VQE applications. 

\begin{figure}[t]
    \centering
    \includegraphics[width=.23\textwidth]{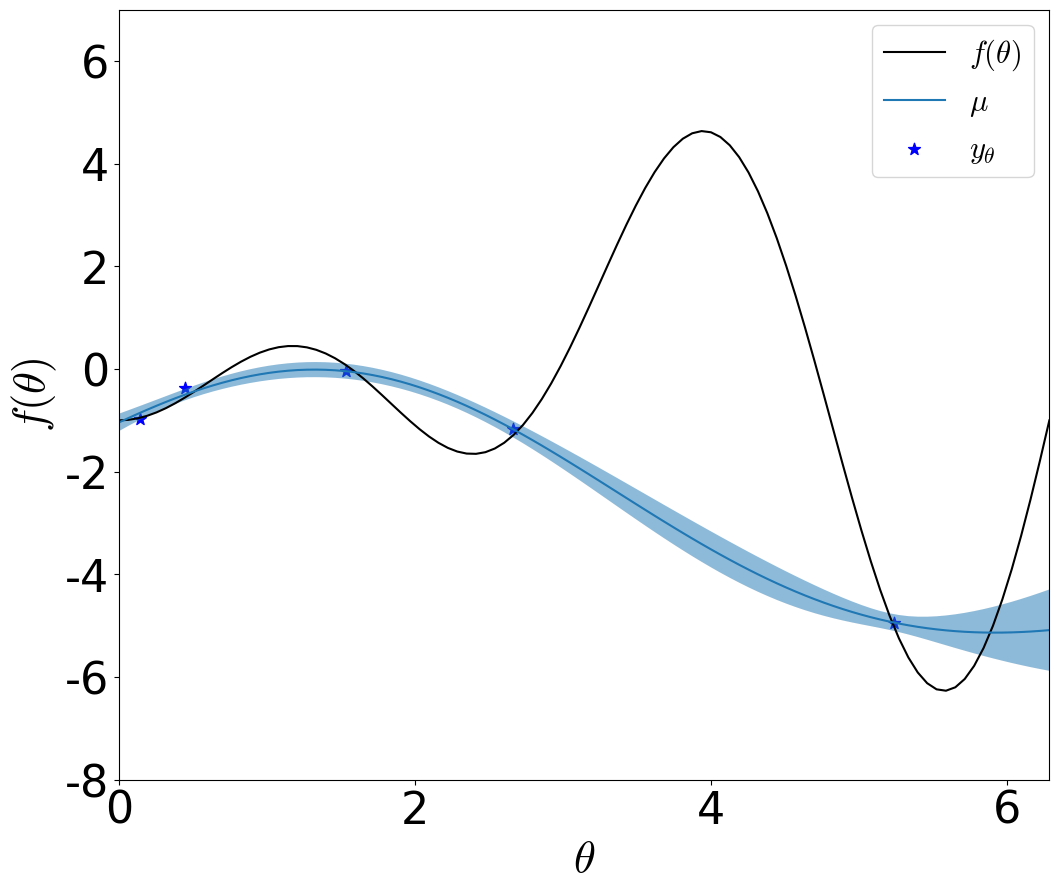}
    \includegraphics[width=.23\textwidth]{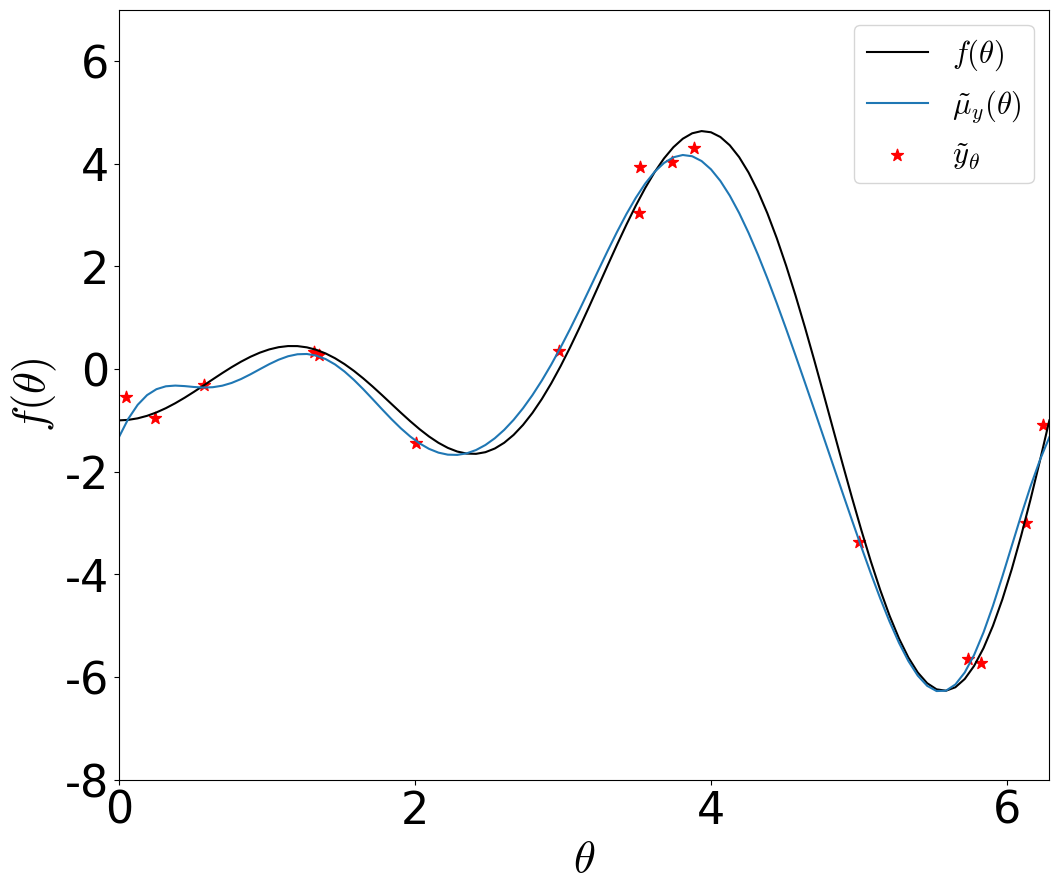}
    \includegraphics[width=.23\textwidth]{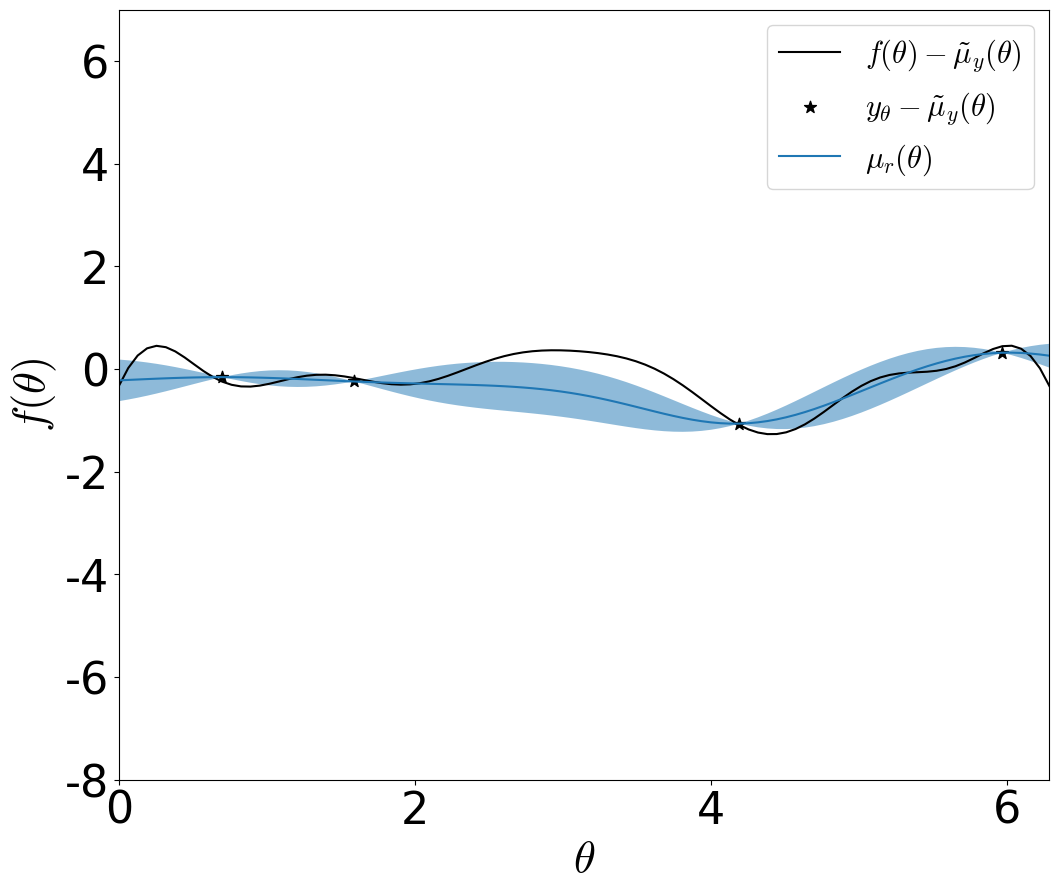}
    \includegraphics[width=.23\textwidth]{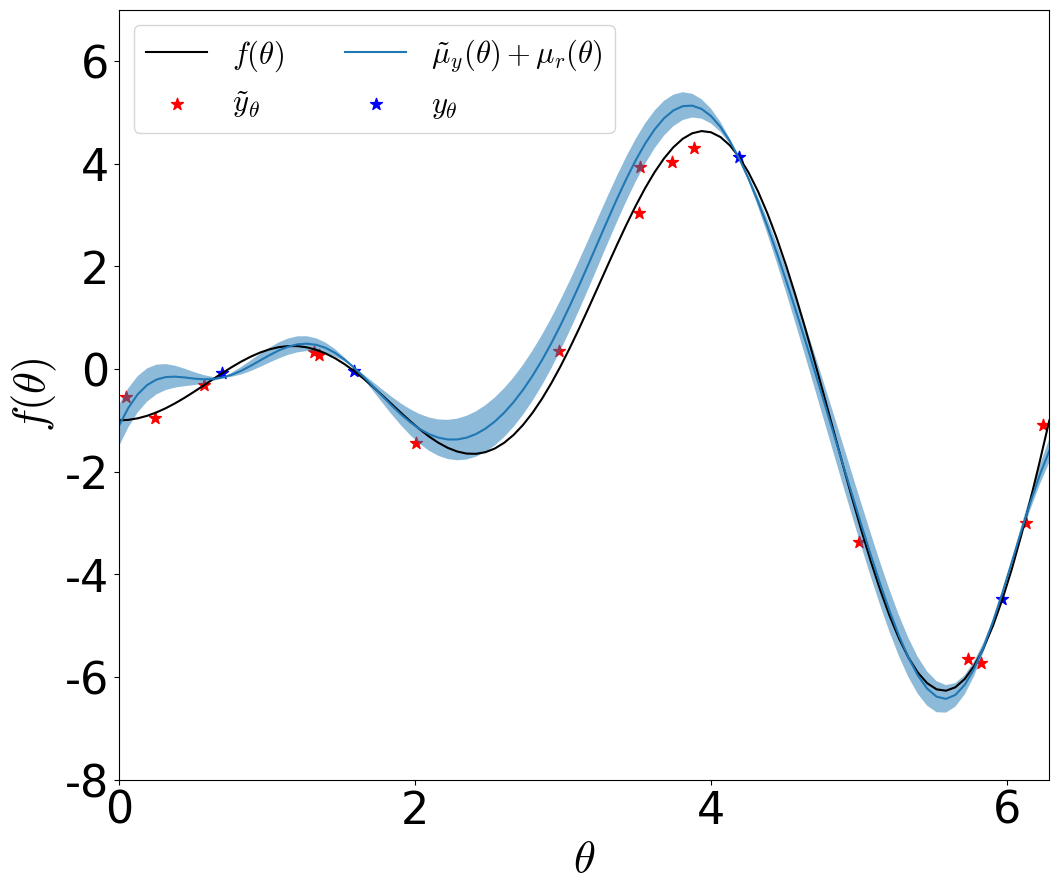}
    \caption{BOPT illustrations: (Left) Standard GP modeling approach, fit directly to 5 high-shot observations. (Center left) the topological prior was constructed from 16 noisy observations. (Center right) GP model of the residual fit using only 4 high-shot observations. (Right) the GP with topological prior of the objective function using the same budget spent on constructing the standard GP shown in the leftmost plot.   }
    \label{fig:motivation}
\end{figure}

We present an illustration of the motivation behind BOPT in Figure \ref{fig:motivation}, where we use additive normal noise terms, with 1 high-shot observation having the same cost as 16 low-shot observations (i.e., $\epsilon_{S}\sim\mathcal{N}(0,0.1^2)$, and $\epsilon_{s}\sim\mathcal{N}(0,0.4^2)$).
In the leftmost plot, a standard GP is trained on 5 high-shot samples; the model is poorly fit due to the insufficient volume of data.
On the other hand, the center-left plot shows that 16 noisy observations can be used to learn an accurate model of the surface topology $\tilde{\mu}_y(\theta)$, providing a better fit than the standard GP model with just 20\% of the budget expended.
The center-right plot shows the true residual $f(\theta) - \tilde{\mu}_y(\theta)$ and its corresponding GP model using 4 high-shot observations $y_\theta$.
The rightmost plot represents the complete GP-TP model of the true objective function $f(\theta)$, which uses the same budget spent on constructing the standard GP shown. 

BOPT can take advantage of any standard BO acquisition function by simply replacing the standard GP model with the proposed GP-TP model, whose predictive posterior distribution is denoted by $f(\theta) | \mathcal{D}_n, \text{TP} \sim \mathcal{N}( \mu_{n,\text{TP}}(\theta), \sigma^2_{n,\text{TP}}(\theta) )$. For example, LCB can represented as follows
\begin{equation}\label{eq:bopt_acq}
\alpha_{\text{BOPT-LCB}, n}(\theta) = \mu_{n,\text{TP}}(\theta) - \sqrt{\beta} \sigma_{n,\text{TP}}(\theta),
\end{equation}
To extend EI to the BOPT setting, we replace the analytical solution in \eqref{eq:expected-improvement} with the so-called noisy EI \cite{nei} approach that can more effectively handle noise in the observations. Noisy EI performs an expectation over \eqref{eq:ei_criterion} using values drawn from the predictive posterior distribution.
In this work, we focus predominantly on LCB.
We summarize BOPT for the LCB acquisition function in Algorithm \ref{alg:bopt}.

\begin{algorithm}
\caption{Bayesian Optimization with Priors on Surface Topology (BOPT) Algorithm}
\begin{algorithmic}[1]
\State{\textbf{Input:} Circuit $f$; high-shot count $S$; low-shot count $s$; low-shot data $\mathcal{D}^\text{low-shot}_{\text{init}}$ constructed by spending $B_{\text{init}}$ on $\tilde{y}_\theta$; and total budget $B$.}
\State{\textbf{Initialize:} Domain $\Theta = [0,2\pi]^d$; High-shot data $\mathcal{D}^\text{high-shot} \leftarrow \emptyset$; and $k=B_{\text{init}}$.}
\State{Construct surrogate $\tilde{\mu}_y$ using $\mathcal{D}^\text{low-shot}_{\text{init}}$} \label{alg:step:surrogate_based_prior}
\State{Initialize GP-TP model via \eqref{eq:gp-tp-prior}}
\While{$k \leq B$}
\State{Minimize $\alpha_{\text{BOPT-LCB}, k}(\theta)$ to find $\theta_{k+1}$ }
\State{Evaluate quantum circuit $y(\theta_{k+1}, s=S)$}
\State{Update high-shot dataset $\mathcal{D}^{\text{high-shot}}\leftarrow\{ \theta_{i}, y(\theta_{i},S)-\tilde{\mu}_y(\theta_{i}) \}_{i=0}^{k+1}$ }
\State{Update posterior GP-TP model using $\mathcal{D}^{\text{high-shot}}$}
\State{Increment spent budget $k \leftarrow k + 1$}
\EndWhile
\end{algorithmic}
\label{alg:bopt}
\end{algorithm}
\textbf{}

\subsection{Sparse Gaussian Processes}
In the BOPT framework, the algorithm is initialized by taking low-shot samples $\mathcal{D}^\text{low-shot}_{\text{init}}$, which contains a relatively large number of samples, to construct $\tilde{\mu}_y$ appearing in the topological prior.
While this surrogate can be constructed using any data-driven model, it must be capable of handling large data sets containing significant amounts of noise. 
In this work, we are interested in using GPs due to their simple and effective uncertainty quantification properties, particularly when dealing with shot noise in quantum circuits \cite{iannelli2021noisy}.
However, the covariance matrix inverse in equation \eqref{eq:gp_var} results in quadratic memory and cubic computational complexity, making the construction and inference with a GP-TP from large datasets a limiting factor. 
To improve the limit on the number of low-shot observations that can be used, we use sparse GP models \cite{titsias2009svgp} to construct $\tilde{\mu}_y(\theta)$.

Assume that a given dataset $\mathcal{D}^\text{low-shot}_{\text{init}}$ contains $m$ data points and that computational or memory requirements limit us to constructing a GP with at most $\ell$ data points, where $\ell<m$.
To fit the GP model, a set of \emph{inducing points},  $\hat{\mathcal{D}}_{\ell}$ must be learned to minimize the statistical distance between the true GP conditioned on $\mathcal{D}^\text{low-shot}_{\text{init}}$ and an approximate GP conditioned on $\hat{\mathcal{D}}_{\ell}$. 
A common approach is to learn an approximate predictive posterior as a function of a latent distribution over the inducing points $q_\xi(\hat{\Theta})$ with hyperparameters $\xi$, where $\hat{\Theta} = \{\hat{\theta}_i\}_{1}^{\ell}\in \hat{D}_{\ell}$  \cite{titsias2009svgp, Hensman2013svgp, Burt2020svgp, moss2023svgp}, known as a sparse variational
\footnote{The term \emph{variational} in machine learning parlance denotes a framework to approximate an intractable posterior distribution over the latent variables, with a more tractable distribution. This is a different usage of \emph{variational} in quantum algorithms, which denotes a class of methods used to approximate the ground state of a quantum system.} 
GP (SVGP). The SVGP learns $q_\xi(\hat{\Theta})$ by maximizing the evidence lower bound objective (ELBO) function,  
\begin{equation}
   \argmax_{\gamma, \xi, \hat{\Theta}} \operatorname{ELBO}(\gamma, \xi, \hat{\Theta}) = 
    \sum_{i=1}^m 
    q_\xi\left(\hat{\Theta}\right)\ln p_\gamma\left(\tilde{y}_{\theta_i} \mid \hat{\Theta},\theta_i \right) - \operatorname{KL}\left(q_\xi\left(\hat{\Theta}\right) \| p_\gamma\left(\hat{\Theta}\right) \right).
\end{equation}
Here, $\operatorname{KL}(\cdot)$ denotes the Kullback-Leibler divergence between the approximate model and the true posterior $p_\gamma (\hat{\Theta} )$.
The ELBO function does not require the functional form of the true posterior to optimize the approximate model, as the intractable posterior is computed by decomposing Bayes’ rule.
The summation in the ELBO is based on the assumption that the training data consists of i.i.d. samples (which is generally the case for VQE problems).

For this work, we use the stochastic variational inference (SVI) strategy \cite{Hensman2013svgp}, which provides better scaling by using an explicit representation of the inducing variables. GP regression can be performed using natural gradients \cite{NaturalGradients} of the ELBO with respect to the variational parameters to learn the inducing variables through mini-batched stochastic gradient descent. The SVI method iteratively proceeds by selecting a mini-batch to compute the natural gradients, which can be used to individually update the latent variable distributions $\bigl\{q_\xi\left(\hat{\theta_i}\right) \bigr\}_{i=1}^m$ using inference over $q_\xi\left(\Theta\right)$ where $\Theta = \{\theta_i\}_{i=1}^{m}$.   
By maximizing the ELBO loss function to find the hyperparameters $\xi$ and inducing points $\hat{\Theta}$, we can construct a sparse GP model to approximate the posterior distribution to scale more favorably in terms of the amount of training data.
This SVI approach to constructing the sparse GP provides the mean function that serves as the surrogate-based prior in step \ref{alg:step:surrogate_based_prior} of algorithm \eqref{alg:bopt}.

\section{Results and Discussion}
\label{sec:casestudies}

In this paper, we investigate the performance of our proposed BOPT framework in solving VQE problems.
We consider two molecular systems, $H_2$ and $H_3^+$, and separate the case studies based on the method of quantum simulation.
Specifically, we consider a state vector simulator (SVS), a fake backend, and a physical quantum computer.
The SVS is a classical quantum simulation strategy that allows the simulation of the system without hardware noise by fully determining the state vector $| \Psi \rangle,$ executed in the circuit.
This method would represent the ideal quantum computer, but the cost of simulating such a circuit grows exponentially with the number of qubits, limiting its usage to small systems.
The fake backend simulations perform the SVS simulations using a data-driven noise model that resembles specific quantum computers, allowing for detailed performance analysis on noisy simulators. These noise models allow the simulation of various transpilation strategies, which map the circuit to a specific quantum processor topology. 
The fake backends used in this work are based on IBM's quantum computers, Cairo, Kolkata, and Mumbai. All of them implement the Falcon processor layout and use an unoptimized transpilation.
These fake backends more closely resemble the behavior of existing quantum hardware, yet making it classically simulatable.
We aimed to corroborate our classical simulations with hardware runs on the same quantum hardware.
Unfortunately, because of the quick turnaround and development of QPUs, IBM made all classically simulated noise models unavailable while drafting this manuscript.
Finally, we execute our algorithm on actual quantum hardware using IBM's Torino quantum computer. This computer uses the Heron processor and is considerably larger than the previously mentioned devices (133 vs. 27 qubits), but a noise model is not yet available for it.

In both SVS and noisy simulators, we use the Hartree-Fock initialization with a HEA using $R_y$ gates and entangling CNOT gates. We consider ansatz depths of 4 layers, and the chemistry Hamiltonian is constructed on the STO-3G basis. 
The $H_2$ problem is represented with 4 qubits (16 parameters), and the $H_3^+$ problem is represented with 6 qubits (24 parameters).
Optimization runs use a total budget $B =  150$, with the BayesOpt methods allocating $B_{\text{init}} = 30$ for initialization.
The circuits are sampled using a high-shot count $\bar{s} = 100,000$, except when initializing BOPT methods, which use a low-shot count of $\underline{s} = 1,000$. 

In all results, we present our method compared to the Powell optimization algorithm \cite{powell1964efficient}, which has been identified as a well-balanced optimization algorithm for VQE problems \cite{Singh23}.
Our proposed algorithm, uses the GP-TP and the LCB acquisition function, denoted by \emph{BOPT}, while using the EI acquisition function is denoted as \emph{BOPT (EI)}. 
In addition, we used several variants of the BayesOpt algorithm to ablate the proposed modifications. The proposed method using a Mat\'ern 2.5 is denoted \emph{BOPT (Mat\'ern)} (or \emph{BOPT (Mat\'ern, EI)}).
The standard GP models with a Mat\'ern 2.5 is denoted by the choice of acquisition function (\emph{EI} or \emph{LCB}), while methods that use the periodic kernel are denoted by \emph{EI-p} or \emph{LCB-p}.

\subsection{State Vector Simulators}

\begin{figure}
    \centering
    \includegraphics[width=.3\textwidth]{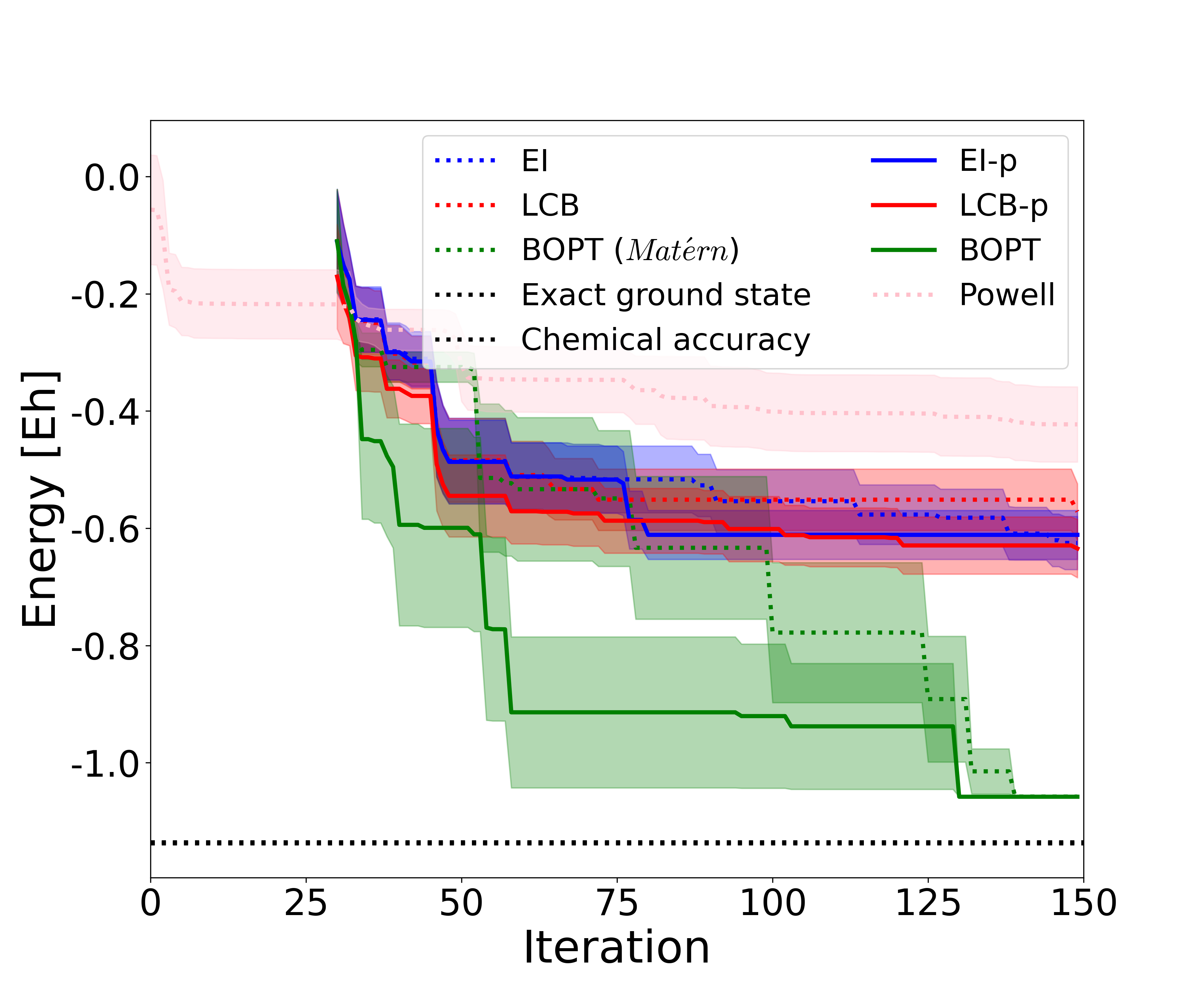}
    \includegraphics[width=.3\textwidth]{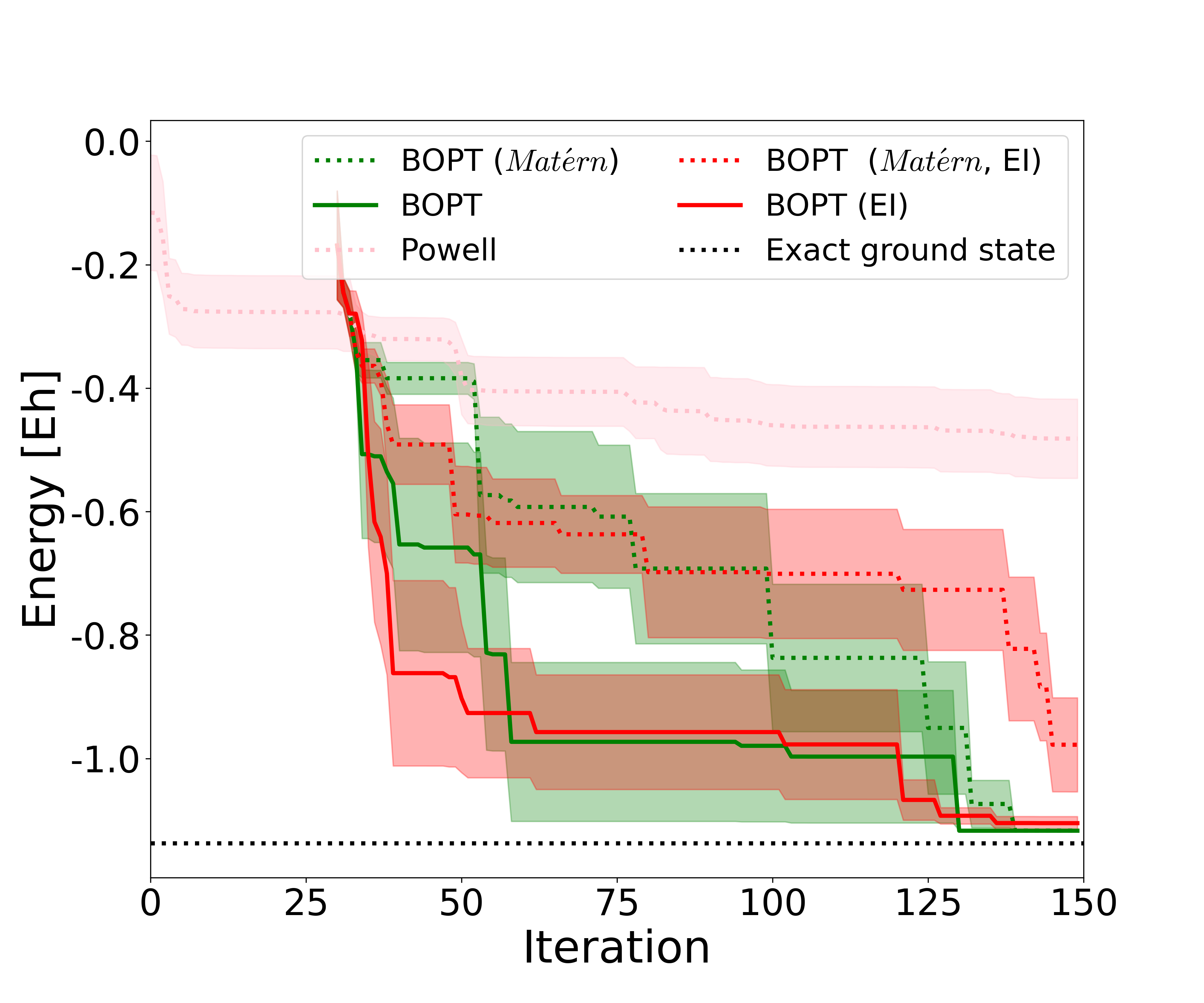}
    \includegraphics[width=.3\textwidth]{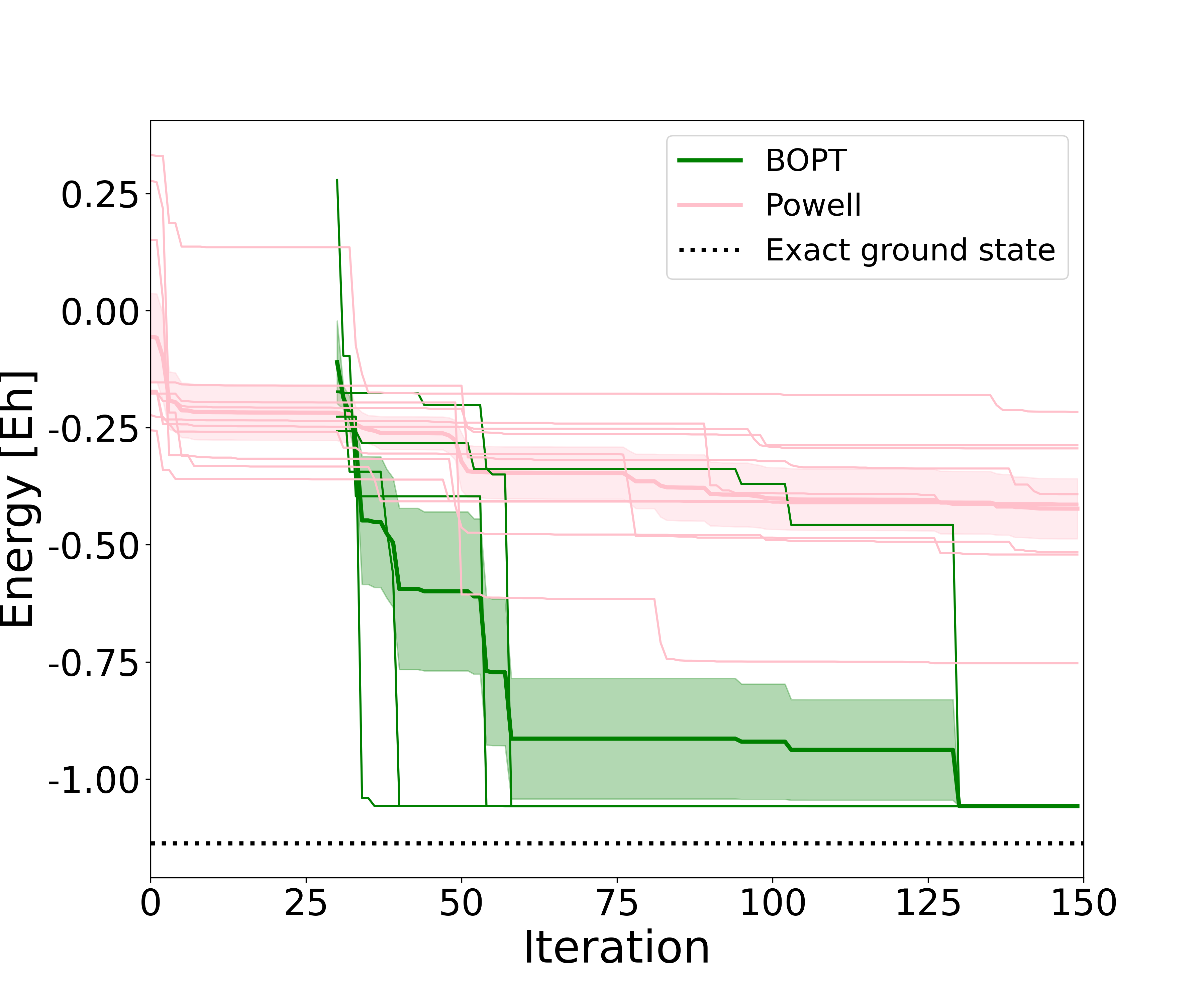}
    \\
    \includegraphics[width=.3\textwidth]{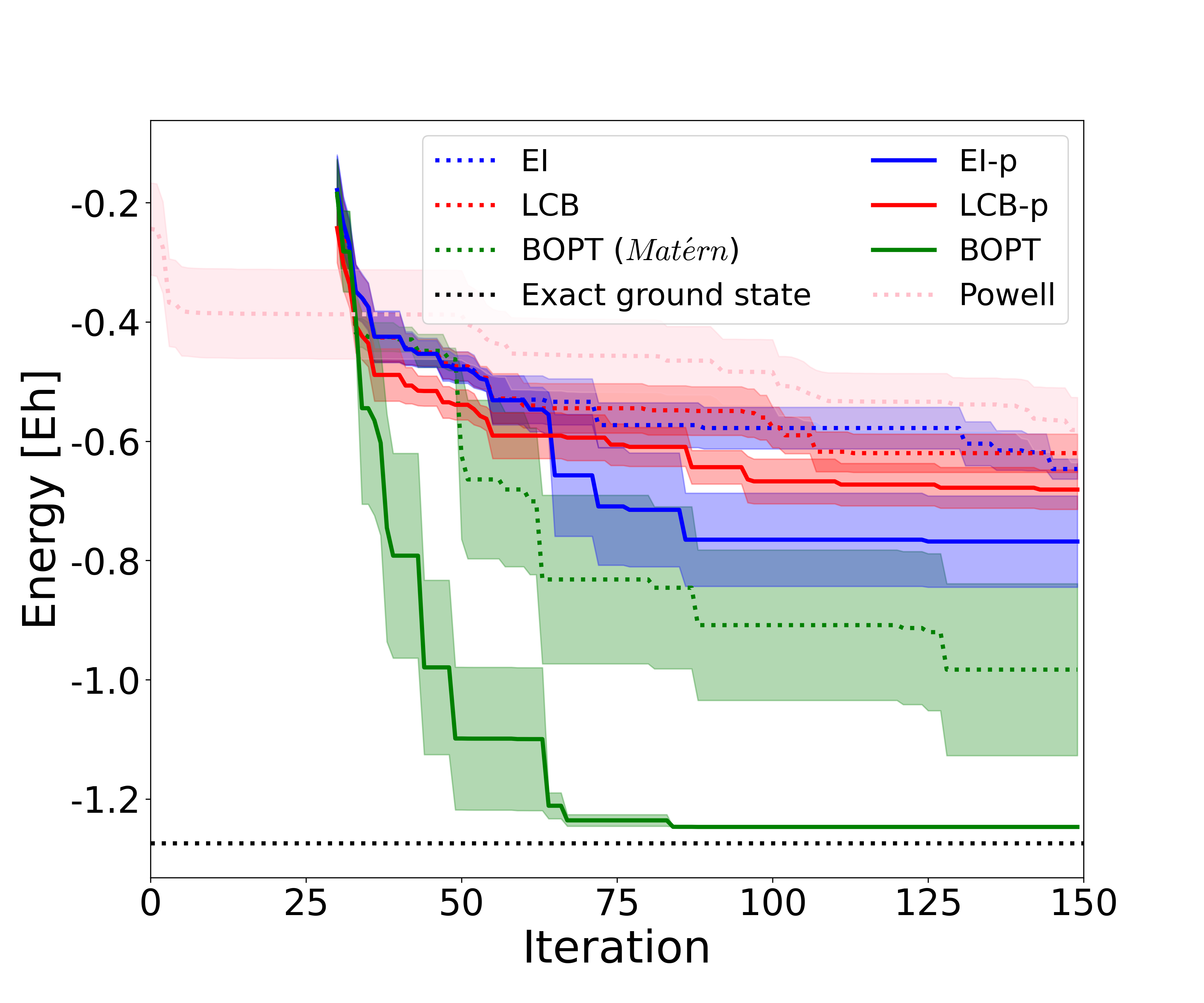}
    \includegraphics[width=.3\textwidth]{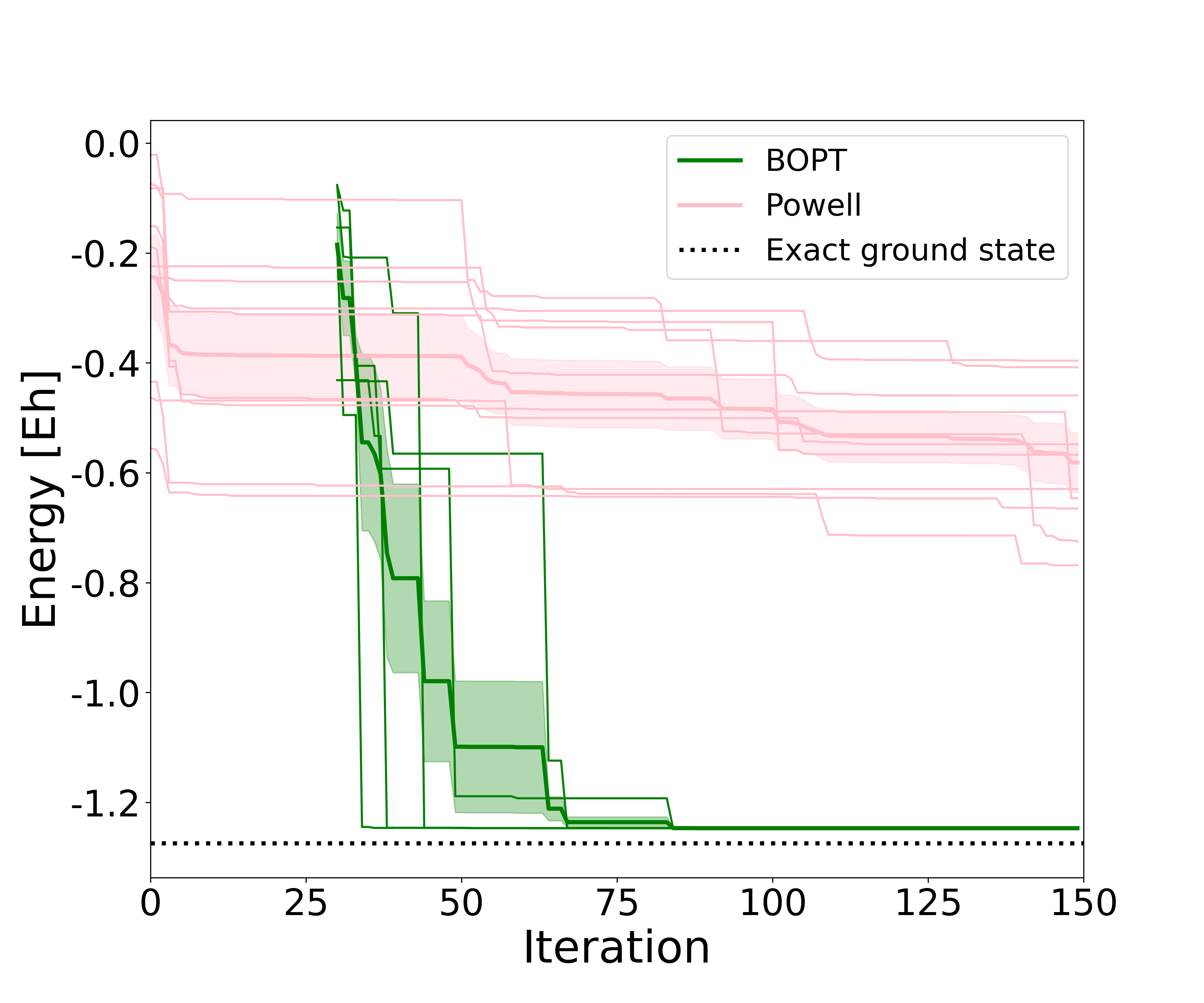}
    \caption{(top) $H_2$,  (bottom) $H_3^+$. (left) Ablation of the periodic kernel and topological prior benchmarked against Powell. (center) Comparison of BOPT methods with EI and LCB acquisition function. (right) Individual runs for the BOPT with periodic kernel relative to Powell. All convergence curves in this figure are generated from 10 runs and show the standard error in the shaded region.     }
    \label{fig:svs_simulations}
\end{figure}

We present the convergence results for the SVS-based VQE in Figure \ref{fig:svs_simulations}.
The upper row shows the results for $H_2$, and the lower row for $H_3^+$.
The leftmost column compares Powell, BOPT and BOPT (Mat\'ern), and the standard BayesOpt for both EI and LCB acquisition functions.
The plot shows that even standard BayesOpt converges notably faster than Powell, with the fastest convergence achieved by the BOPT methods.
Furthermore, BOPT (Mat\'ern) and LCB-p/EI-p methods show that both the periodic kernel and the topological prior have a significant effect on the convergence rate. However, the kernel effects without the topological prior are less pronounced on the $H_2$ problem relative to the $H_3^+$ problem.
A possible explanation for this behavior is that $H_2$ has fewer parameters but the same number of samples, resulting in a more dense set of samples relative to $H_3^+$.
If there are sufficient samples, the periodicity of the parameter space can be learned from observations without the explicit prior, and thus, it would not be expected to accelerate convergence.
In contrast, the less dense set of samples over the $H_3^+$ parameter space may be insufficient to identify periodic behavior, thus benefiting from the explicit inclusion of this prior.

In the center column of Figure \ref{fig:svs_simulations}, the two BOPT methods (with and without the periodic kernel) are presented for each acquisition function. The results show that both acquisition functions perform comparably, with the EI finding slightly higher energy on $H_2$ with slightly faster convergence.
These results, coupled with those from the first column, suggest that the choice of acquisition function does not impact performance as much as choices affecting the model.
On the other hand, the BOPT framework provides a clear advantage over the standard GP model, and the periodic kernel provides an advantage (particularly when used in combination with BOPT) over a standard Mat\'ern 2.5 kernel.        
 
In the third column of Figure \ref{fig:svs_simulations}, the individual runs by Powell and BOPT are shown. These results show that Powell, which uses a sample-based approximation of the gradients, requires many samples before finding a direction of improvement. Conversely, BOPT rapidly converges in the early iterations, likely due to exploiting the topological prior, but appears to have a tendency to get ``stuck," preventing further improvement.               
The exploration-exploitation trade-off allows BayesOpt to find near-optimal solutions quickly. However, in some cases, it may over-explore once it finds a near-optimal solution, which results in behavior similar to that observed in the BOPT method. This pathology has been observed in other VQE problems and addressed by \cite{Muller22}, where the authors propose using a local optimization strategy initialized from the best solution found by BayesOpt after expending a given sample budget.

\subsection{Noisy Simulators}

\begin{figure}
    \centering
    \includegraphics[width=.3\textwidth]{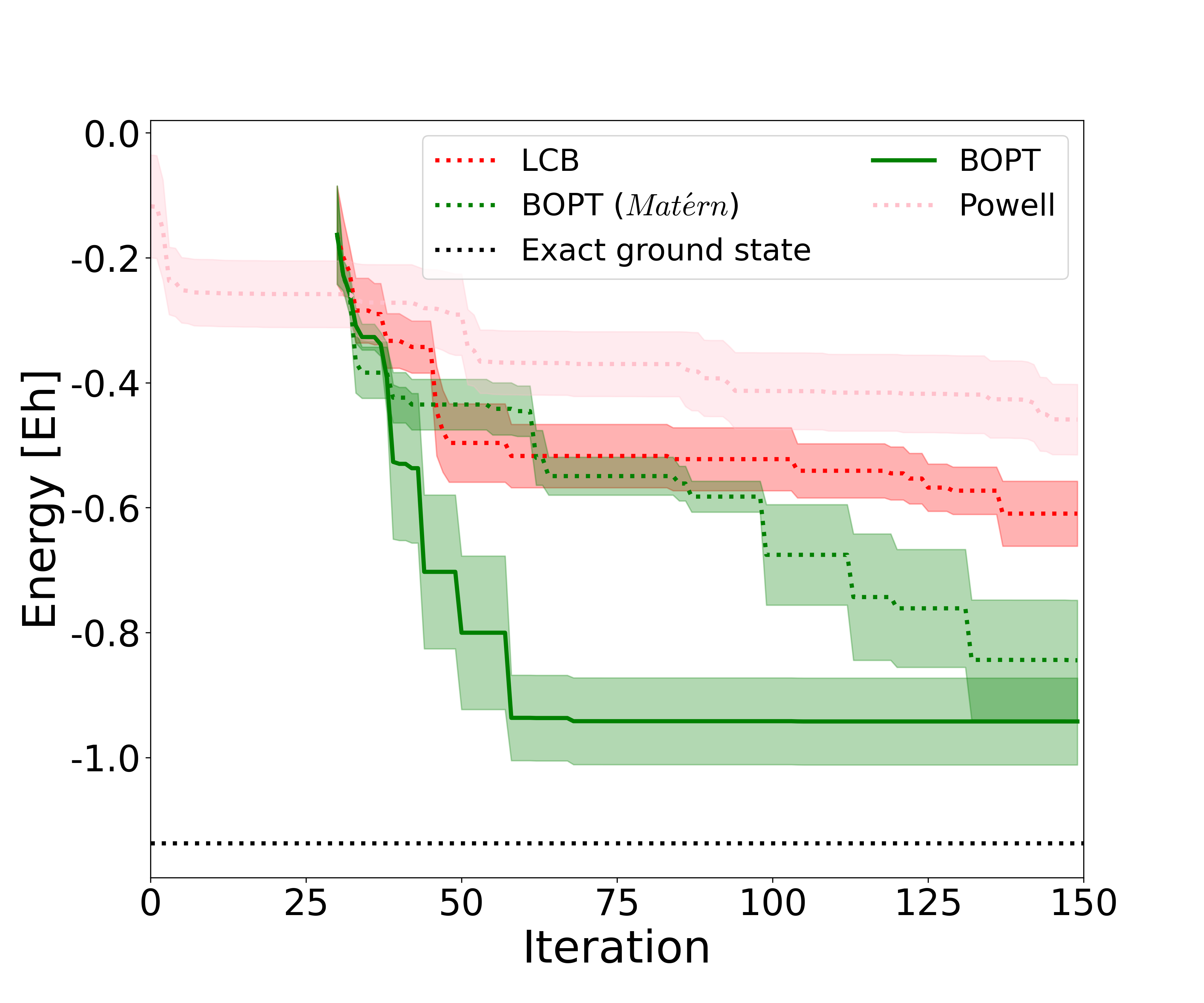}
    \includegraphics[width=.3\textwidth]{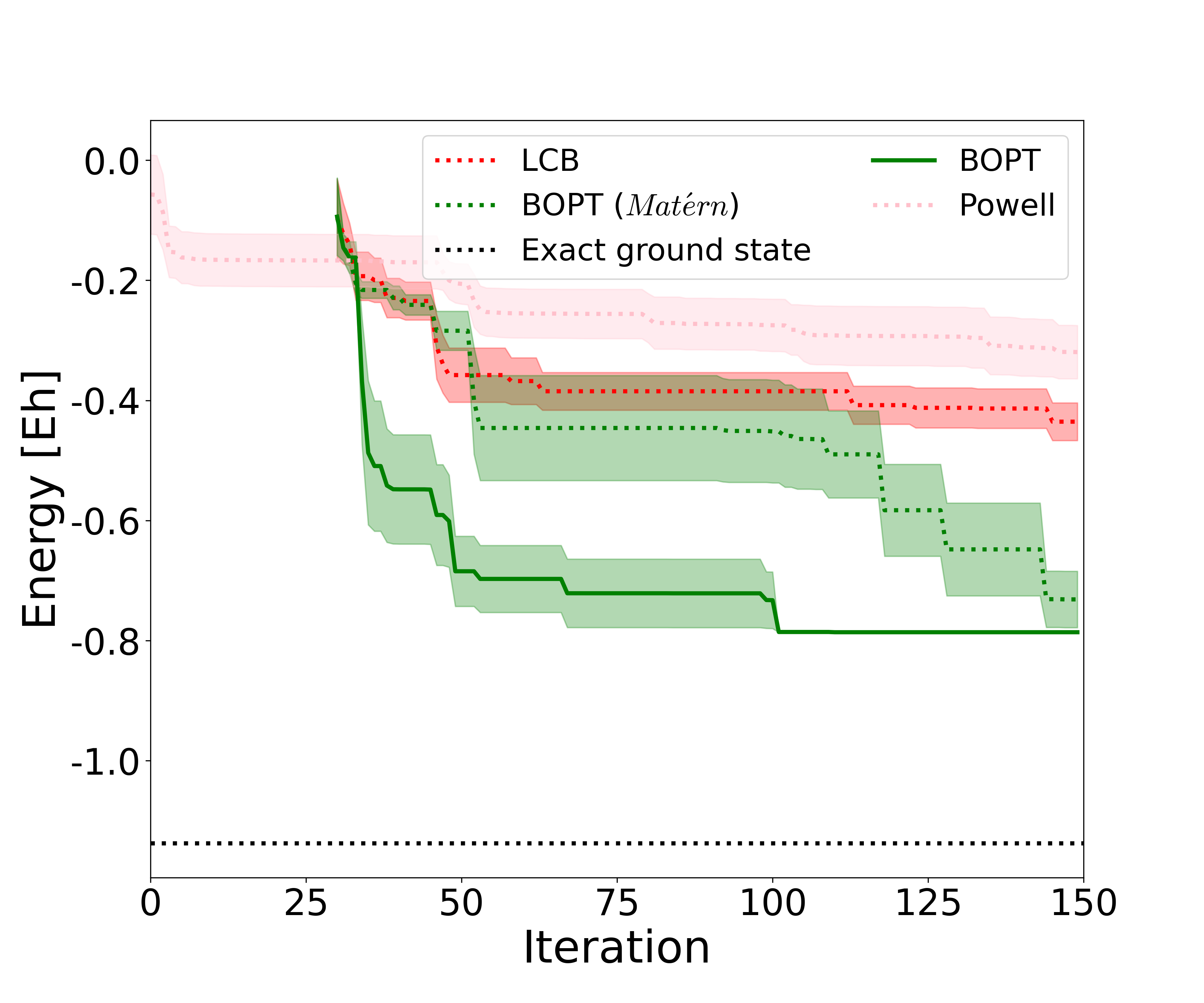}
    \includegraphics[width=.3\textwidth]{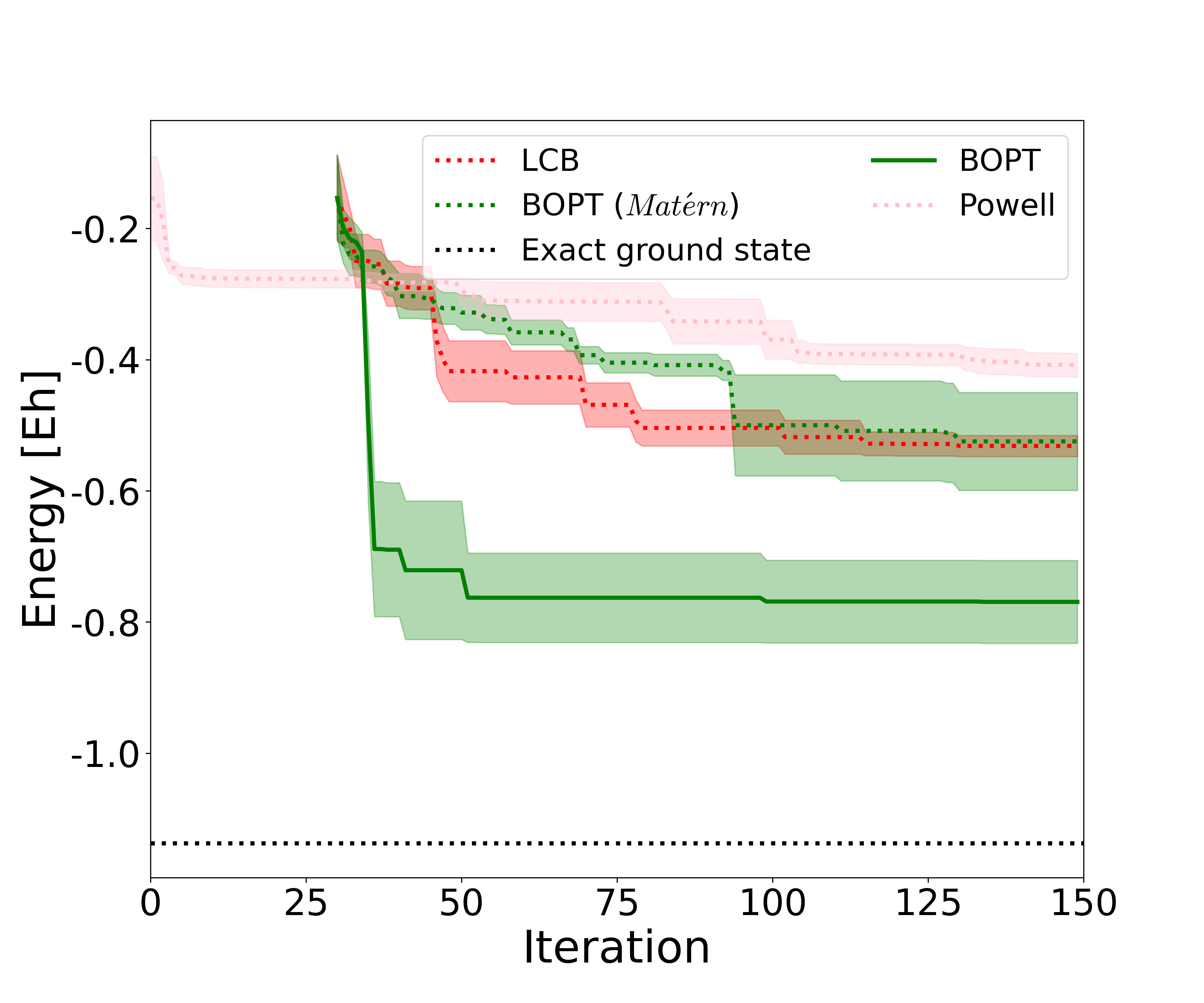}
\\
    \includegraphics[width=.3\textwidth]{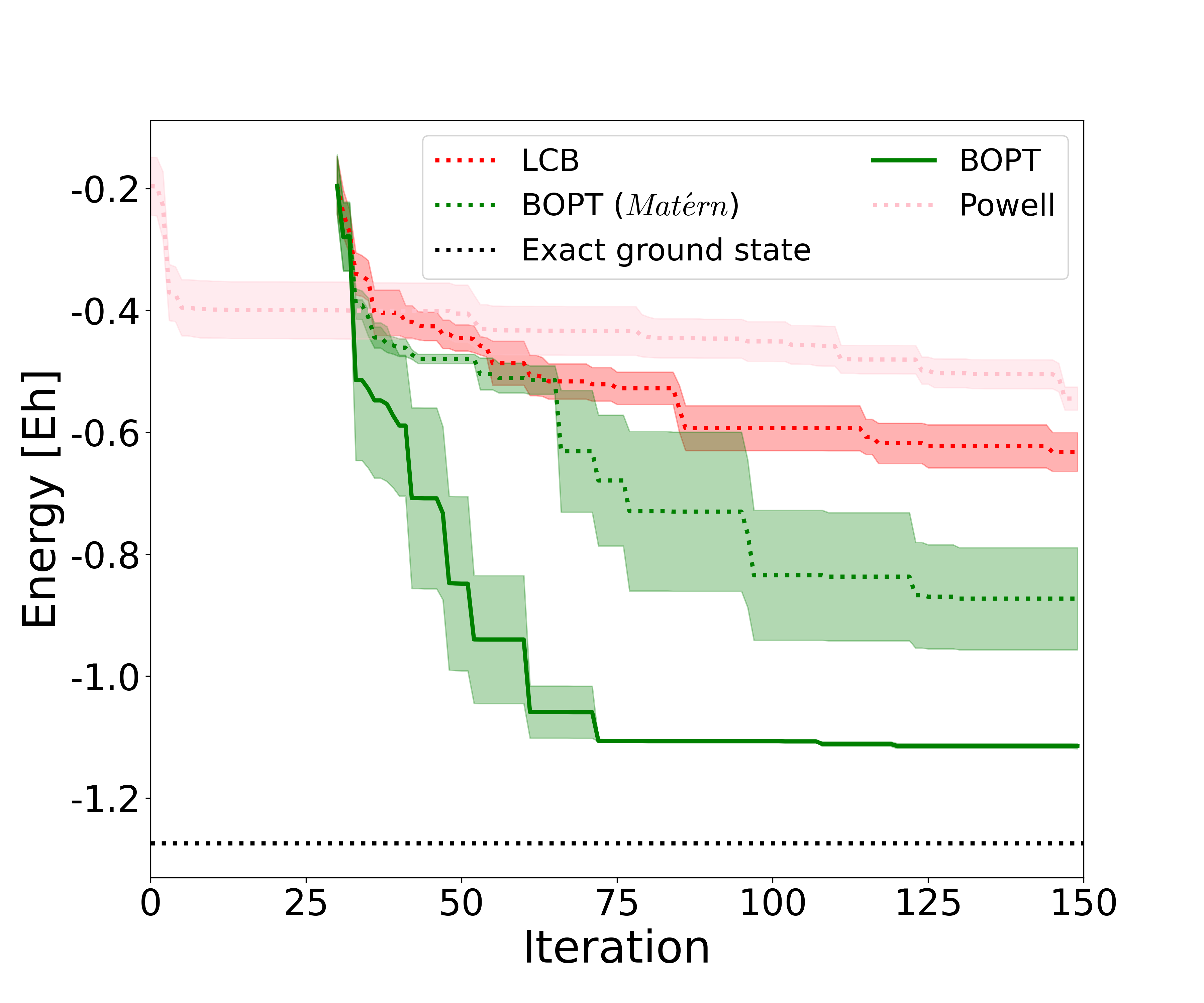}
    \includegraphics[width=.3\textwidth]{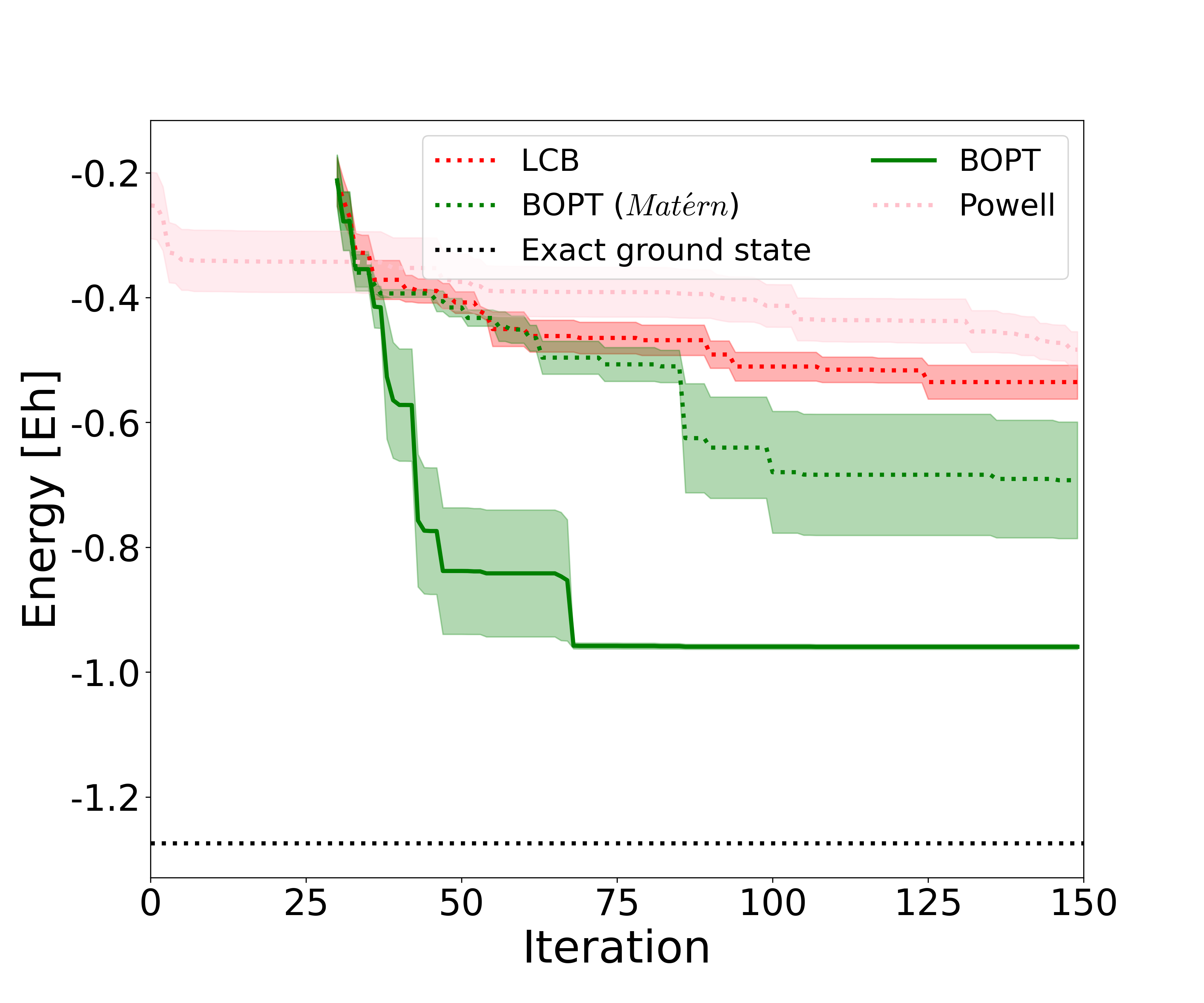}
    \includegraphics[width=.3\textwidth]{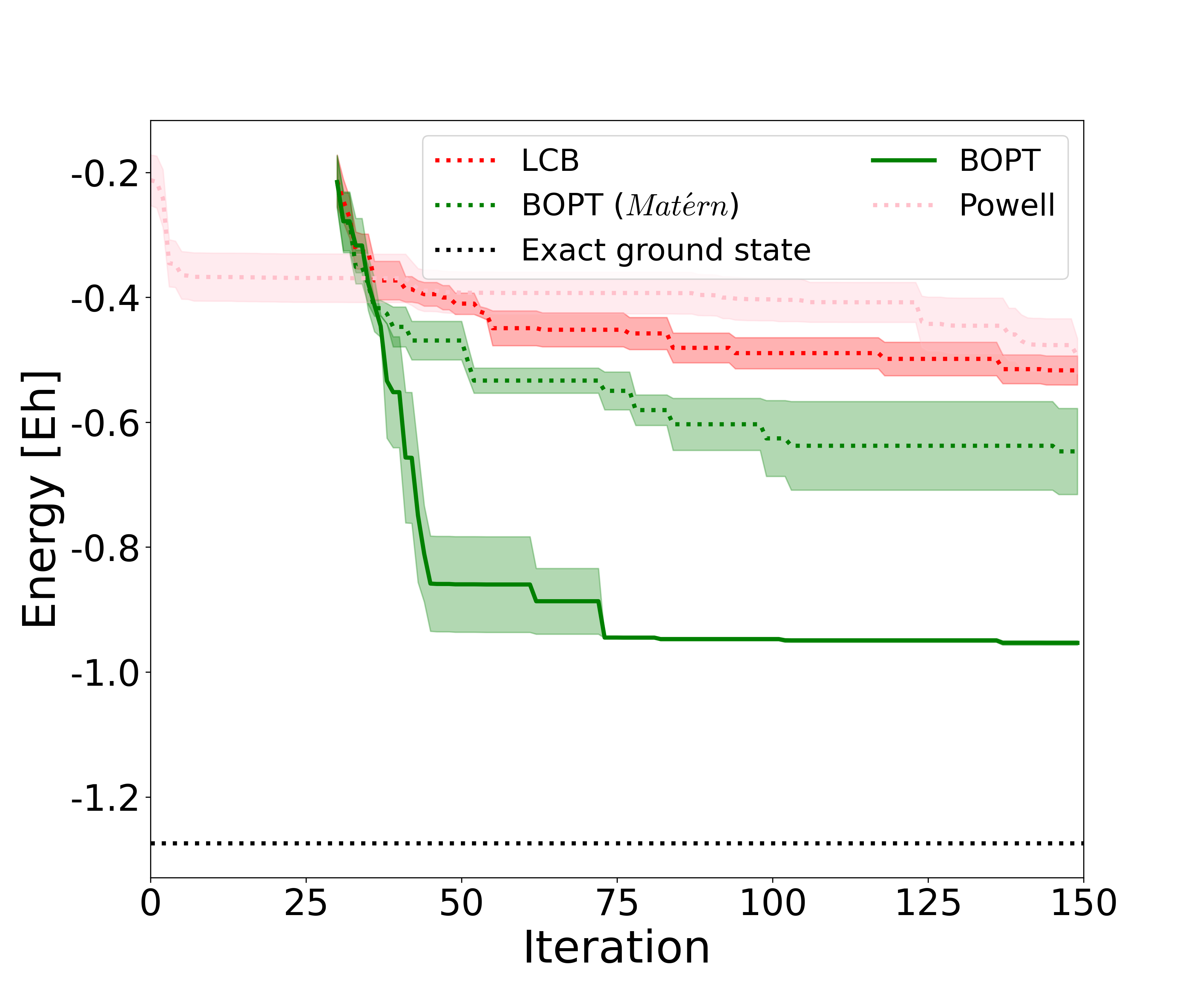}
    \caption{(top) $H_2$, (bottom) $H_3$; convergence results for several different optimization frameworks on fake backend (left) Kolkata, (center) Mumbai, and (right) Cairo. }
    \label{fig:noisy_convergence_summary}
\end{figure}

Next, we present the optimization results on using noisy quantum simulators based on IBM's quantum computers Cairo, Kolkata, and Mumbai for the $H_2$ and $H_3^+$.
First, we present the convergence results in  Figure \ref{fig:noisy_convergence_summary}  for Powell, LCB, BOPT (Mat\'ern), and BOPT.
The four methods show similar performance across all three simulators, with the highest average performance achieved using Mumbai.
Additionally, we note that the best-observed energies are worse for all methods relative to the noise-free case.
In contrast, the relative performance across the 4 methods is similar to the SVS, with Powel performing the worst, followed by LCB, BOPT (Mat\'ern), and BOPT performing the best.

\begin{figure}
    \centering
    \includegraphics[width=.3\textwidth]{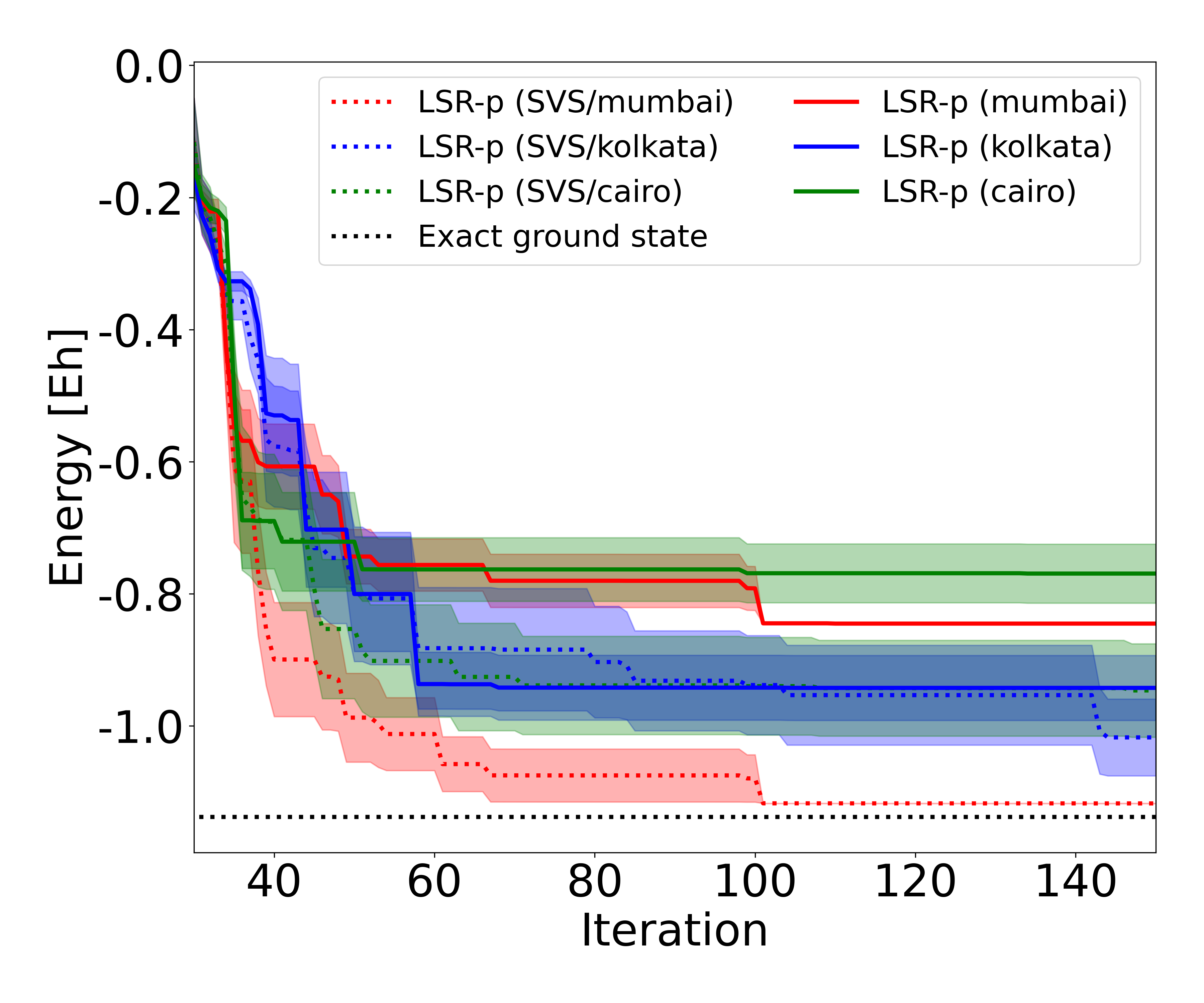}
    \includegraphics[width=.3\textwidth]{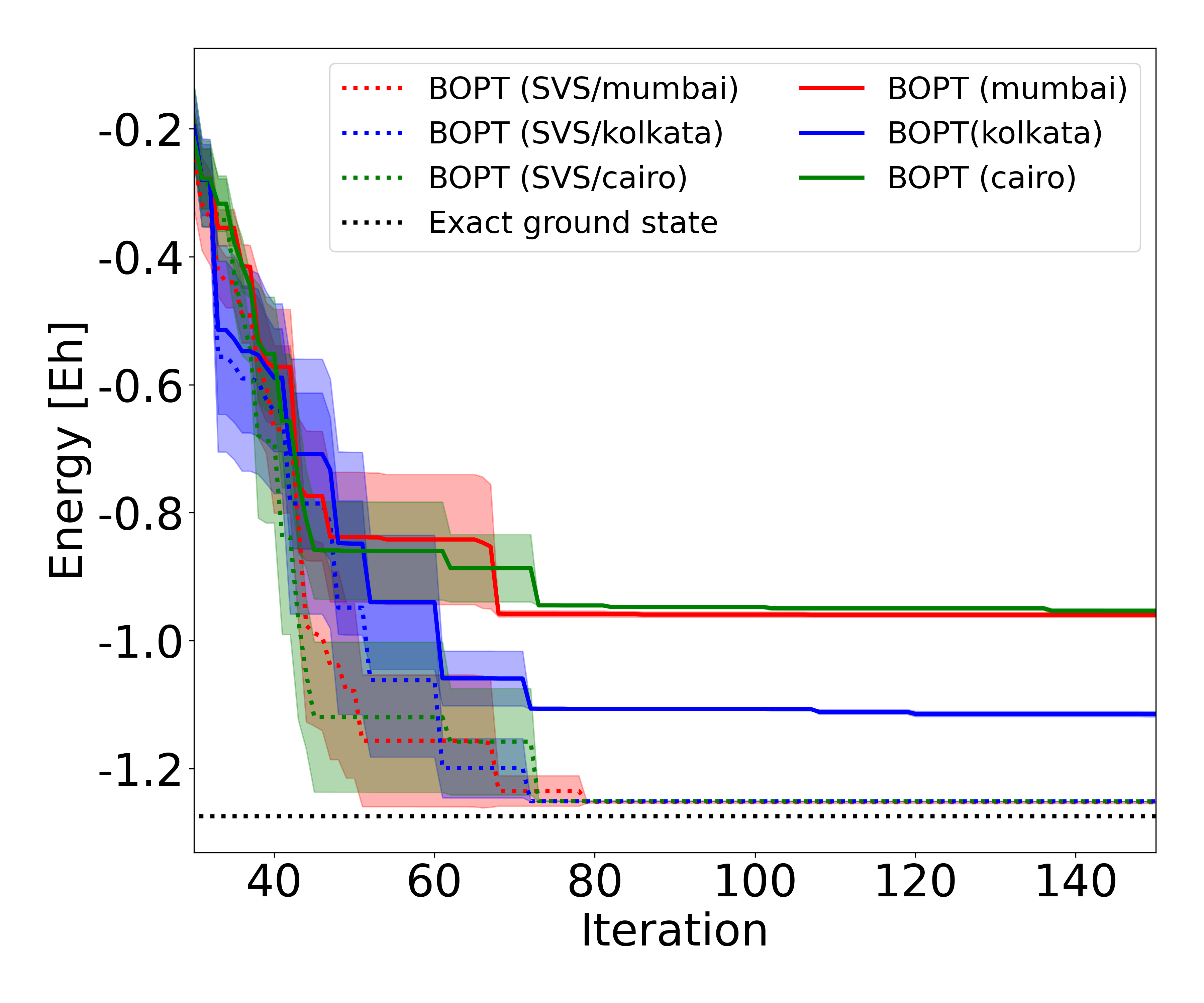}
    \caption{(left) $H_2$, (right) $H_3$; Comparison of the optimization framework on the different QPU models (solid lines), and the energies observed from evaluating the same parameter sequence from optimizing the QPU models on an SVS (dotted lines). The $H_2$ convergence data is generated from 10 runs, and for $H_3$, it is generated from 5 runs. The shaded regions show the standard error.    }
    \label{fig:noisy_convergence_validation}
\end{figure}

On average, the energies observed from the noisy simulators were higher than those observed from the SVS.
However, Figure \ref{fig:noisy_convergence_validation} shows that the sample trace trajectories (the sequence of parameters corresponding to the best-observed energy up to the current iteration), which were generated from optimizing the fake backends, are reevaluated on the SVS.
For all three cases, we see that the actual energies computed from the selected parameters are much lower than what the fake backends report.
Furthermore, we see that the parameters selected by the BOPT framework on both problems are comparable to those generated by optimizing the SVS.
We also note that the behavior of the trace trajectories exhibits similar iteration dynamics, such as an increasing bias for lower energies and similar changes in the standard error. 
Addressing these limitations in accurately measuring the observables from a noisy quantum circuit, yet obtaining the correct convergence toward promising parameters has been observed in other studies \cite{sack2024large} and motivates research in error-mitigation techniques in NISQ devices toward useful application of these devices \cite{kim2023evidence}.

The errors between the fake backends and the validation results from the simulations presented in Figure \ref{fig:noisy_convergence_validation} are presented in Table \ref{tbl:svs_vs_fakebackends} for both problems, using parity plots and histograms.
The histograms show that Kolkata has the smallest noise, while Mumbai and Cairo have similar noise values.
The parity plots show that a bias is present and appears inversely proportional to the magnitude of the observed energy.
We also note that the zero-bias occurs around $-0.2$, which is the average initial value for the optimization methods in Figures \ref{fig:svs_simulations} and  \ref{fig:noisy_convergence_validation}. The parity plots also show a small spread in the energies observed on fake backends at a particular value of the SVS.      

\begin{table}[t]
    \centering
    \caption{Errors between measurements on an SVS compared to the fake backends, using parity plots and histograms of errors between the two models, from data generated from the optimization and validation runs in Figure \ref{fig:noisy_convergence_summary}. }
    \begin{tabular}{|c|c|}\hline
          $H_2$ & $H_3^+$ \\
          \hline
            \includegraphics[width=.23\textwidth]{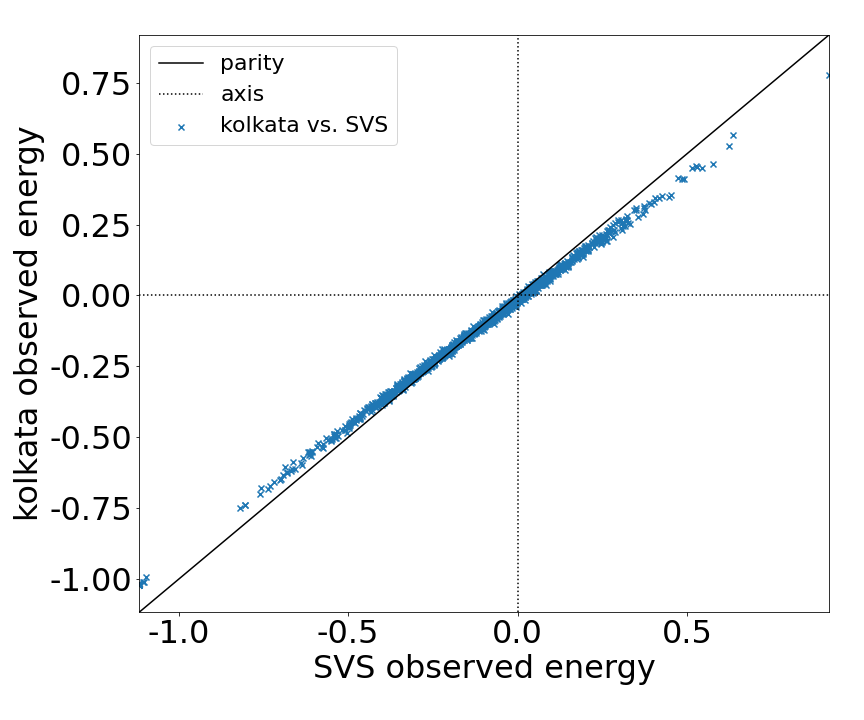}
            \includegraphics[width=.23\textwidth]{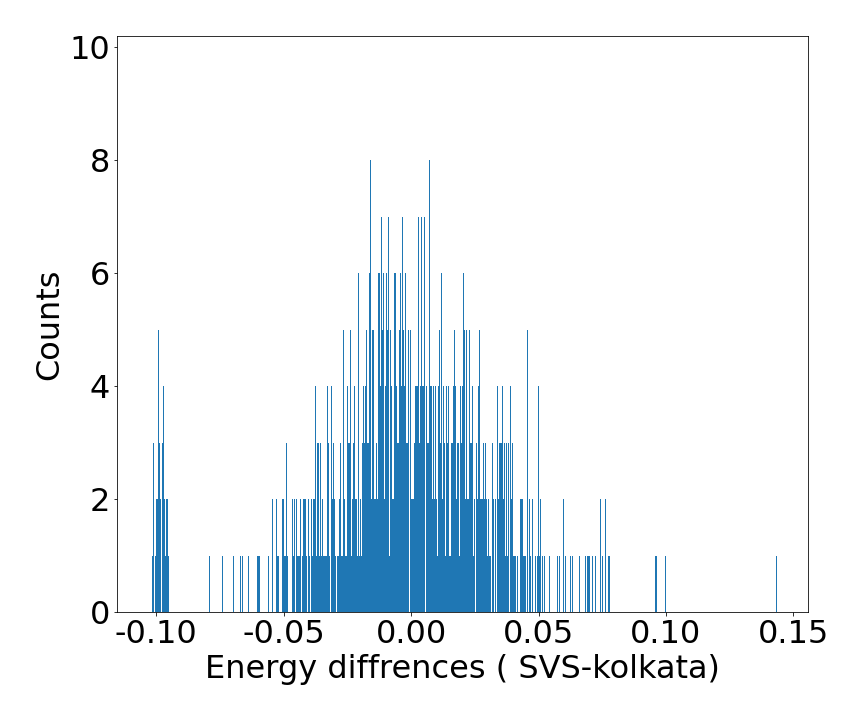}
        &   \includegraphics[width=.23\textwidth]{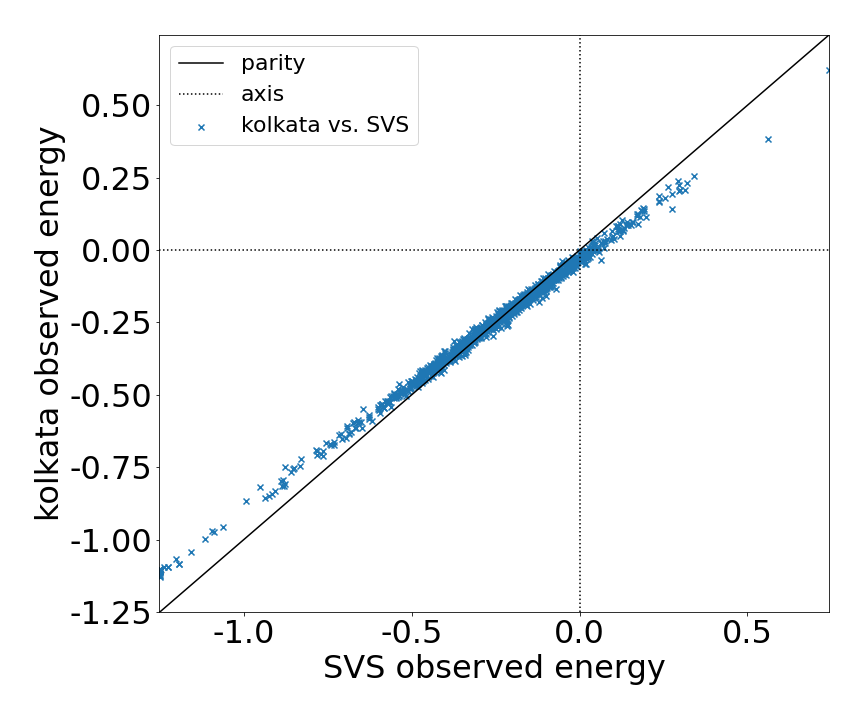}
            \includegraphics[width=.23\textwidth]{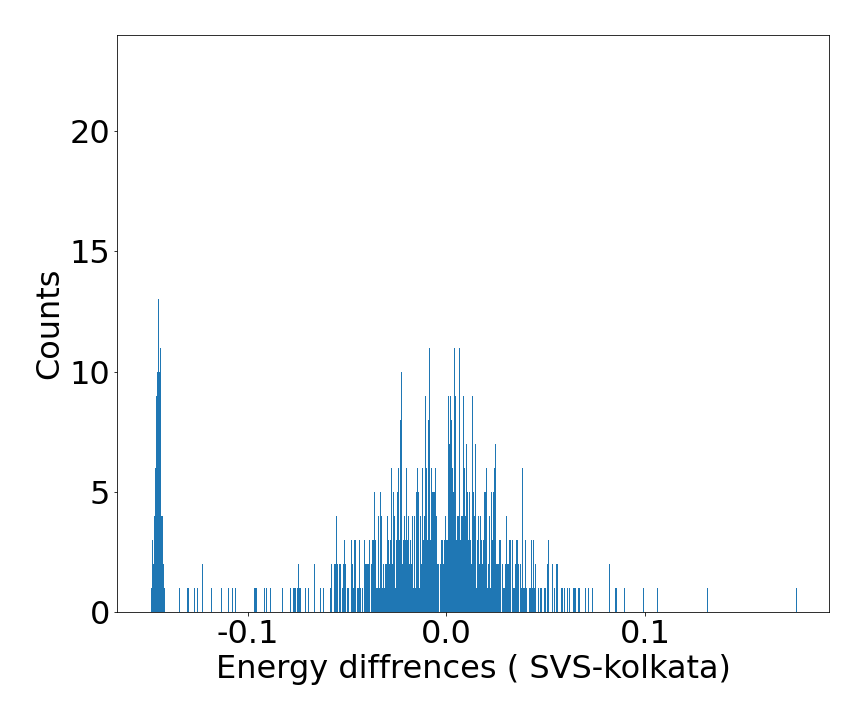} 
\\\hline  \multicolumn{2}{|c|}{Kolkata} \\ \hline
            \includegraphics[width=.23\textwidth]{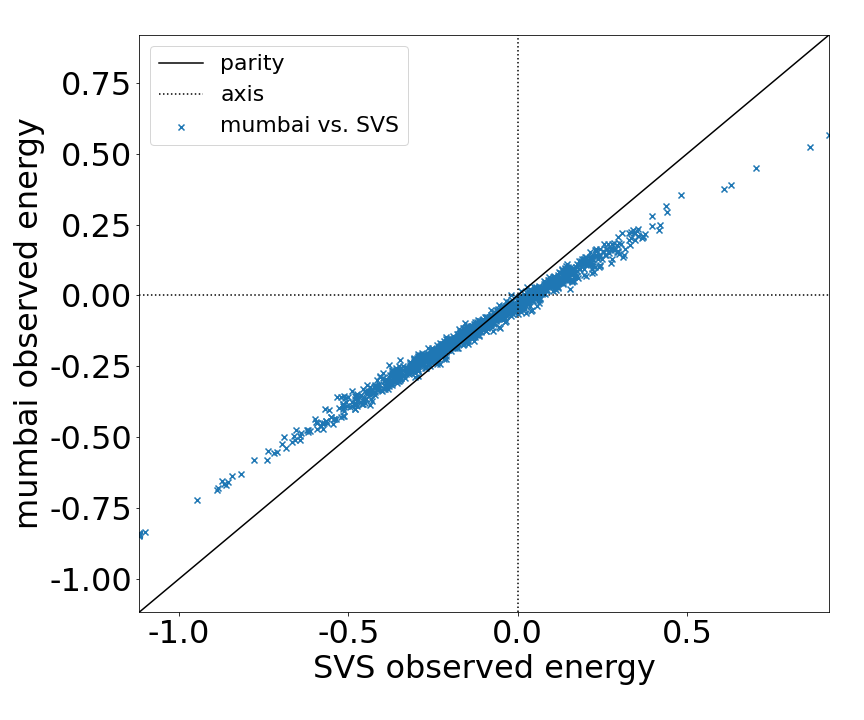}
            \includegraphics[width=.23\textwidth]{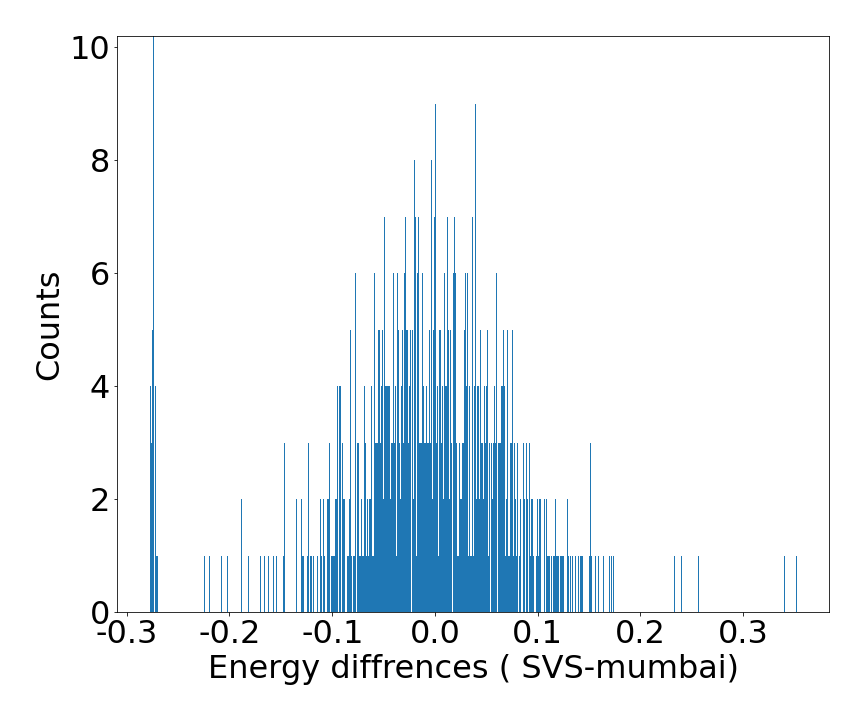}
        &   \includegraphics[width=.23\textwidth]{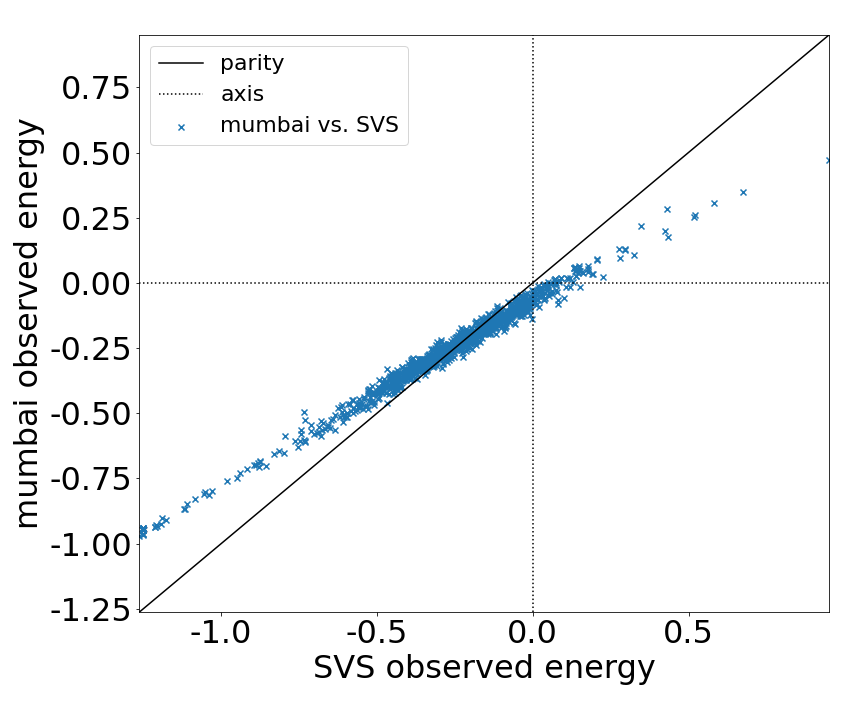}
            \includegraphics[width=.23\textwidth]{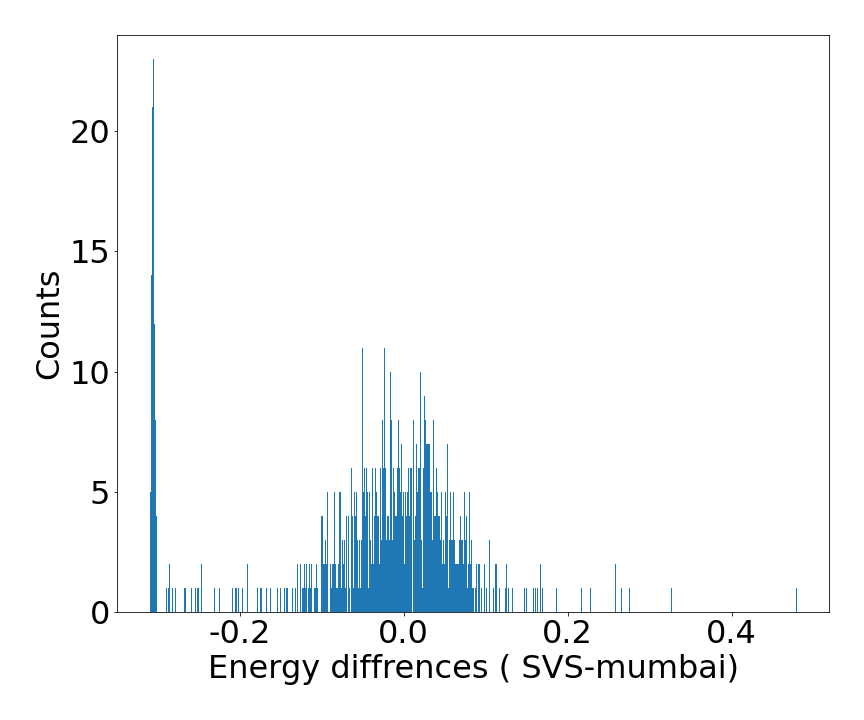}
\\\hline \multicolumn{2}{|c|}{Mumbai}  \\ \hline
            \includegraphics[width=.23\textwidth]{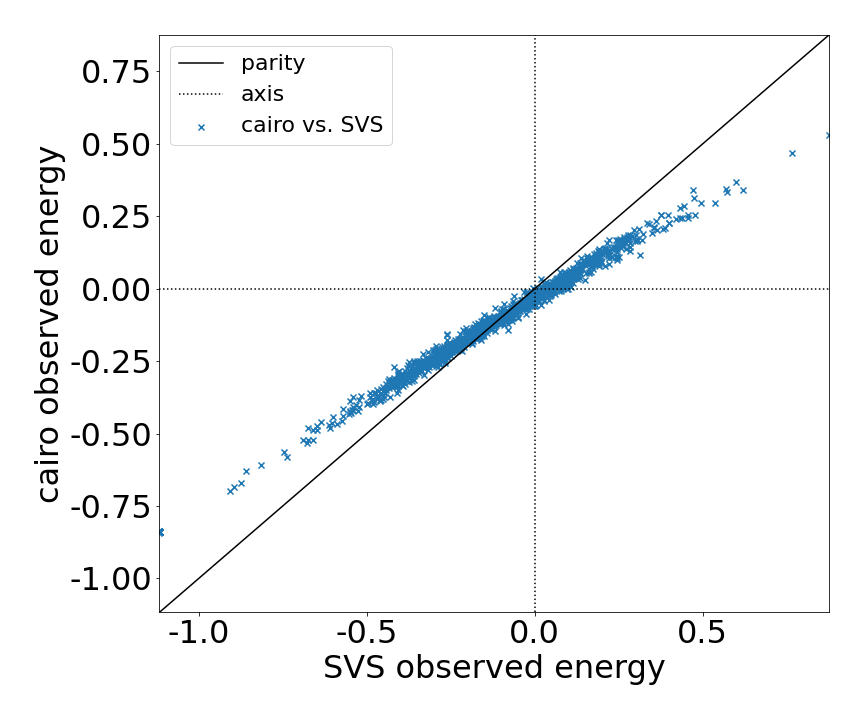}        
            \includegraphics[width=.23\textwidth]{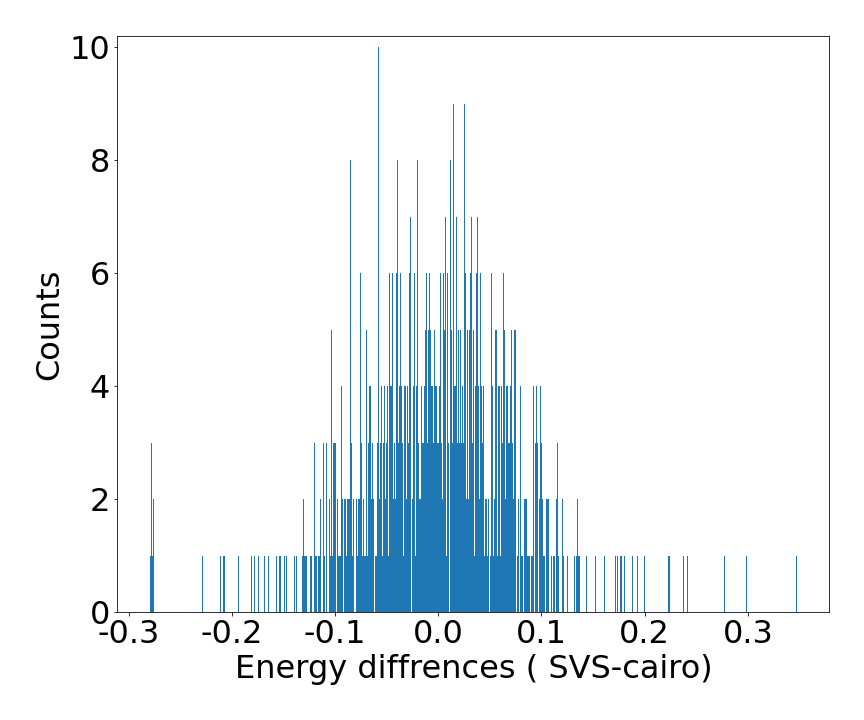}        
        &   \includegraphics[width=.23\textwidth]{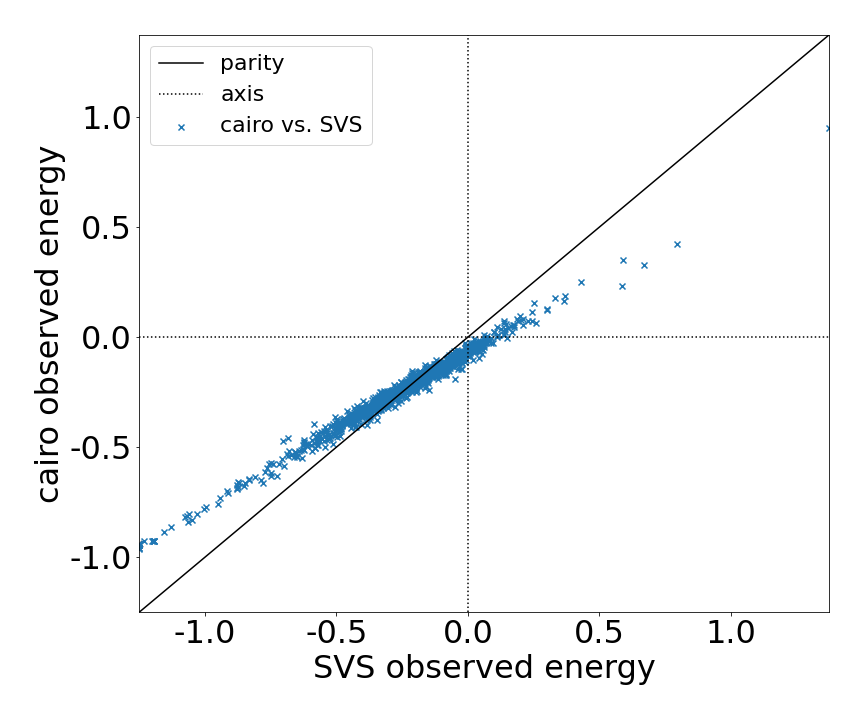}     
            \includegraphics[width=.23\textwidth]{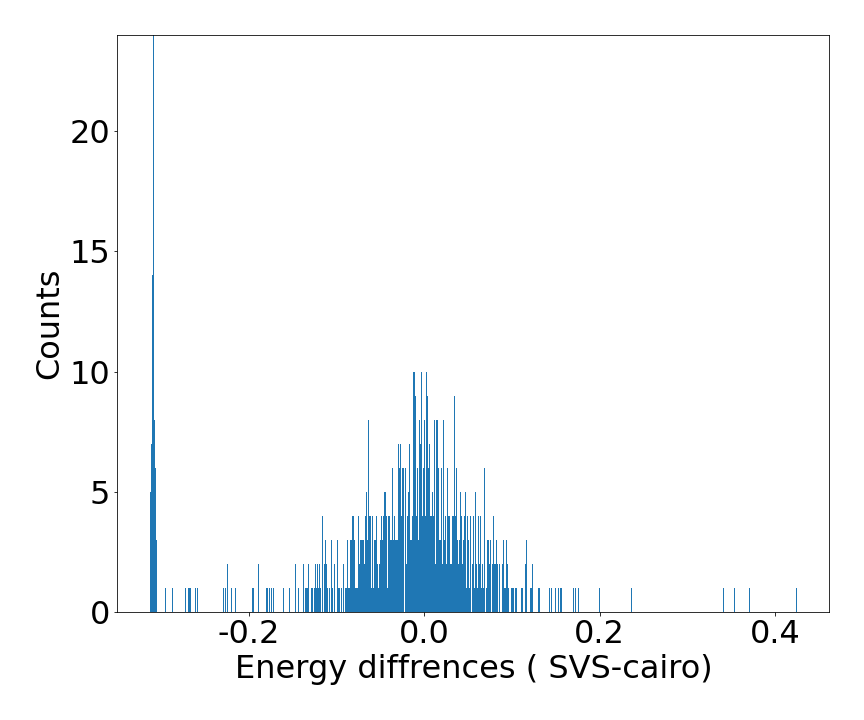}        
\\ \hline \multicolumn{2}{|c|}{Cairo}\\ \hline
    \end{tabular}
    \label{tbl:svs_vs_fakebackends}
\end{table}

These results suggest that the probability distribution representing the hardware noise may have a mean function linear with respect to $\theta$, with a small variance.
These properties of the hardware noise support the use of BayesOpt methods for efficient state preparation on quantum hardware.
The linear bias does not affect the location of the minima, while the small variance allows the GP model to model the response surface accurately using fewer samples.
However, the bias in the hardware noise prevents the usage of current QPUs for determining the ground state energies and requires classical validation, as shown in Figure \ref{fig:noisy_convergence_validation}.

\subsection{Quantum Hardware}

\begin{figure}
    \centering
    \includegraphics[width=0.7\textwidth]{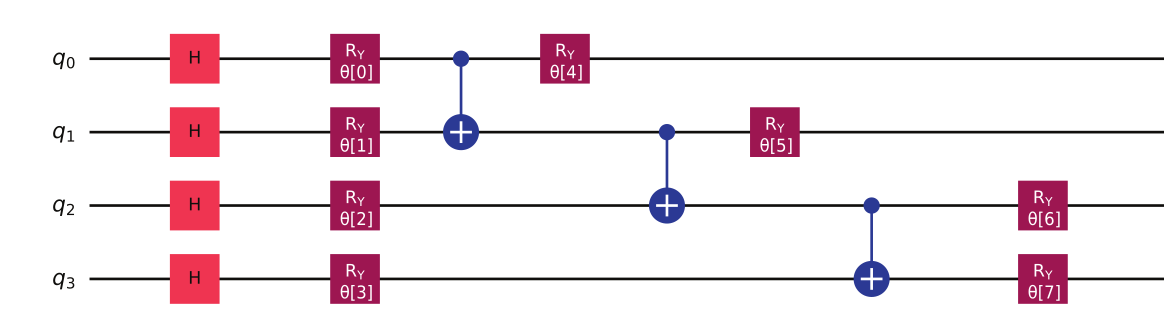}
    \\
    \includegraphics[width=0.9\textwidth]{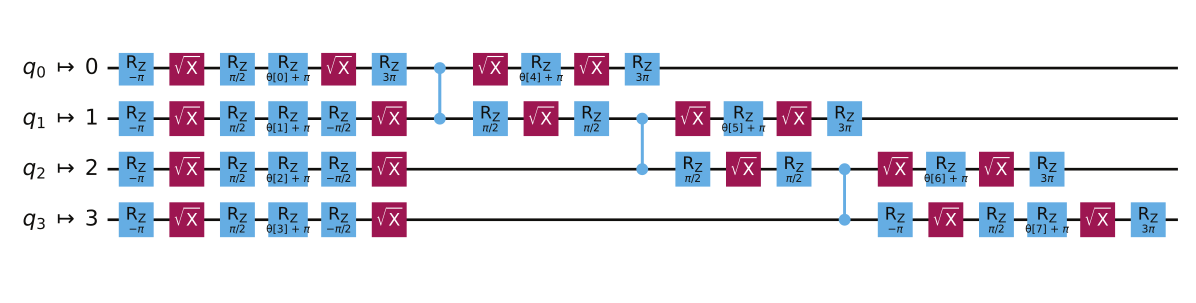}
    
    \caption{The original ansatz circuit (top), and the transpiled circuit (bottom): to run the circuit on IBM Torino, a transpilation step is necessary to express the ansatz only in terms of gates which are natively supported by the QPU (up to a global phase of $3\pi/2$), in this case; $\operatorname{RZ}$, $\operatorname{\sqrt{X}}$, and $\operatorname{CZ}$. It is worth noting that only $\operatorname{\sqrt{X}}$ and $\operatorname{CZ}$ introduce noise, as $\operatorname{RZ}$ is implemented as a noiseless virtual gate on IBM QPUs.}
    
    \label{fig:ansatz_circuits}
\end{figure}

For the last case study, we compare the proposed method with the Powell method to find the ground state of the $H_2$ molecule on physical quantum hardware and validate it on a classical computer.
For the $H_2$ problem, we use an ansatz that restricts the state to real amplitudes, with an initial state of 4 qubits in an equal superposition, $\ket{\Psi_{\text{ref}}} = H^{\otimes 4}\ket{0000}$, where $H^{\otimes 4}$ denotes $4$ Hadamard gates applied on each of the $4$ qubits of the circuit.
The ansatz, shown in Figure \ref{fig:ansatz_circuits} (top), is parameterized by $8$ variables and contains a single entangling layer of $3$ $\operatorname{CNOT}$ gates in a linear layout.
The final circuit that runs on the IBM Torino quantum processor after transpilation is also shown in Figure \ref{fig:ansatz_circuits} (bottom).
Each optimization run is repeated three times, using a total budget $B =  180$, with $B_{\text{init}} = 30$ allocated for initializing the BOPT.
The circuits are sampled using a high-shot count $\bar{s} = 8000$ (the maximum number of shots on IBM Torino being $8192$) and a low-shot count of $\underline{s} = 1000$ for BOPT initialization.

The IBM Quantum job run time for each iteration of the VQEs, comprising $5$ circuits and measurements for the $H_2$ Hamiltonian, was $13$ seconds for both methods.
Function calls on the QPU also incur additional communication time overhead, which includes data transfer, queue, and data retrieval times, as well as pre-processing and post-processing time.
Data transfer time is the time it takes for job data to be sent from the user's local machine to the server and for the job to be created on the server, while data retrieval is the time taken for the results to be sent back to the user's local machine \cite{ibmtime}.
Note that data pre- and post-processing steps might occur concurrently with the communication phases.
Classical pre-processing and IBM server communication time could take up to $45$ seconds per iteration for BOPT and up to $4$ seconds per iteration for Powell, while classical SVS evaluations and optimization took an average of $25$ seconds per iteration for BOPT with $B=$180 and $0.1$ seconds per iteration for Powell, using 8 threads of a 4-core i5-9300H CPU - 2.40GHz processor.
The three BOPT VQEs took more than $6$ hours to run within a single IBM Quantum session, while the three Powell VQEs took almost $5$ hours.
Note that the BOPT run time per iteration is small for the few first iterations, similar to the time taken by Powell, and starts growing as the number of iterations increases because of inference scaling cubically with the number of observations.

\begin{figure}
    \centering
    \includegraphics[width=.49\textwidth]{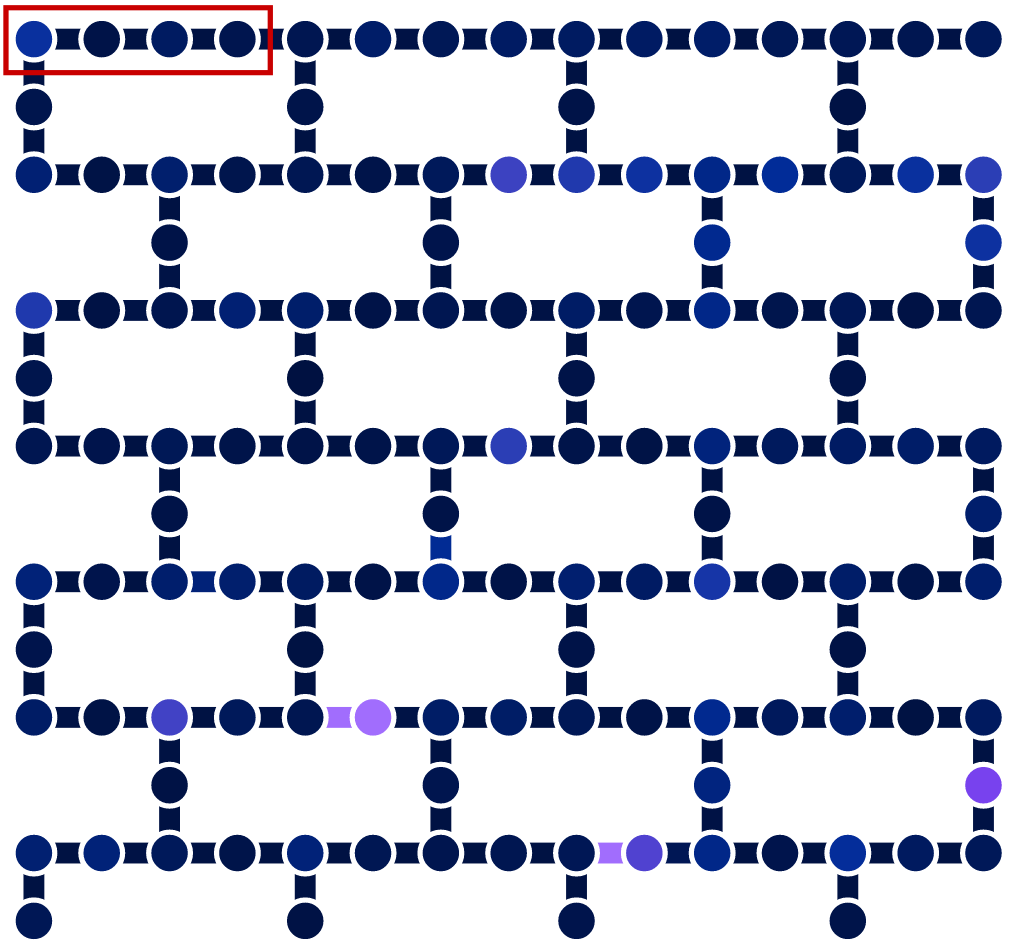}
    \caption{  IBM Torino QPU layout. The  circuit used for $H_2$ ground energy estimation is mapped to the highlighted qubits within the red box.}
    \label{fig:torino_h2}
\end{figure}

\begin{figure}
    \centering
    \includegraphics[width=.45\textwidth]{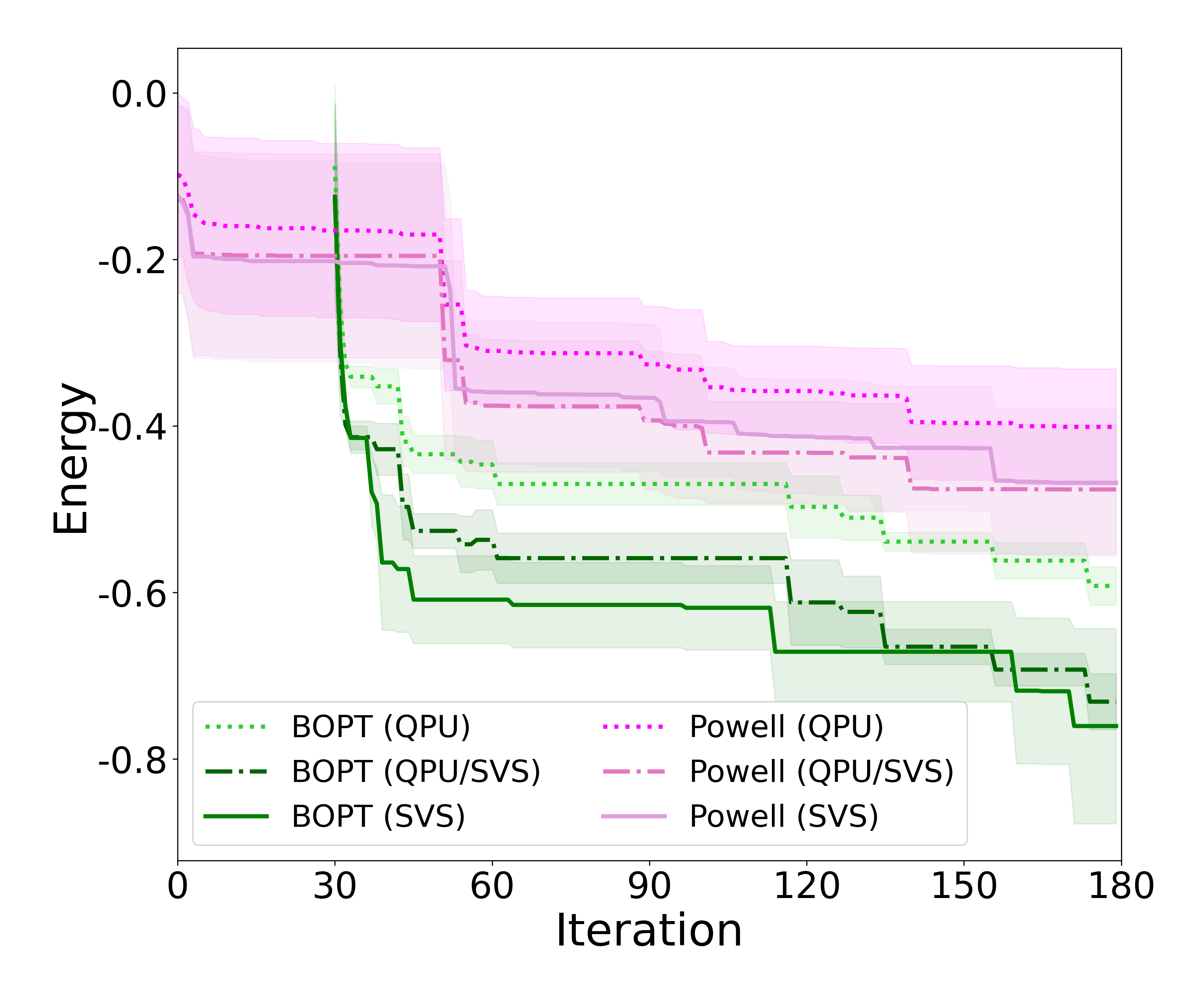}
    \includegraphics[width=.45\textwidth]{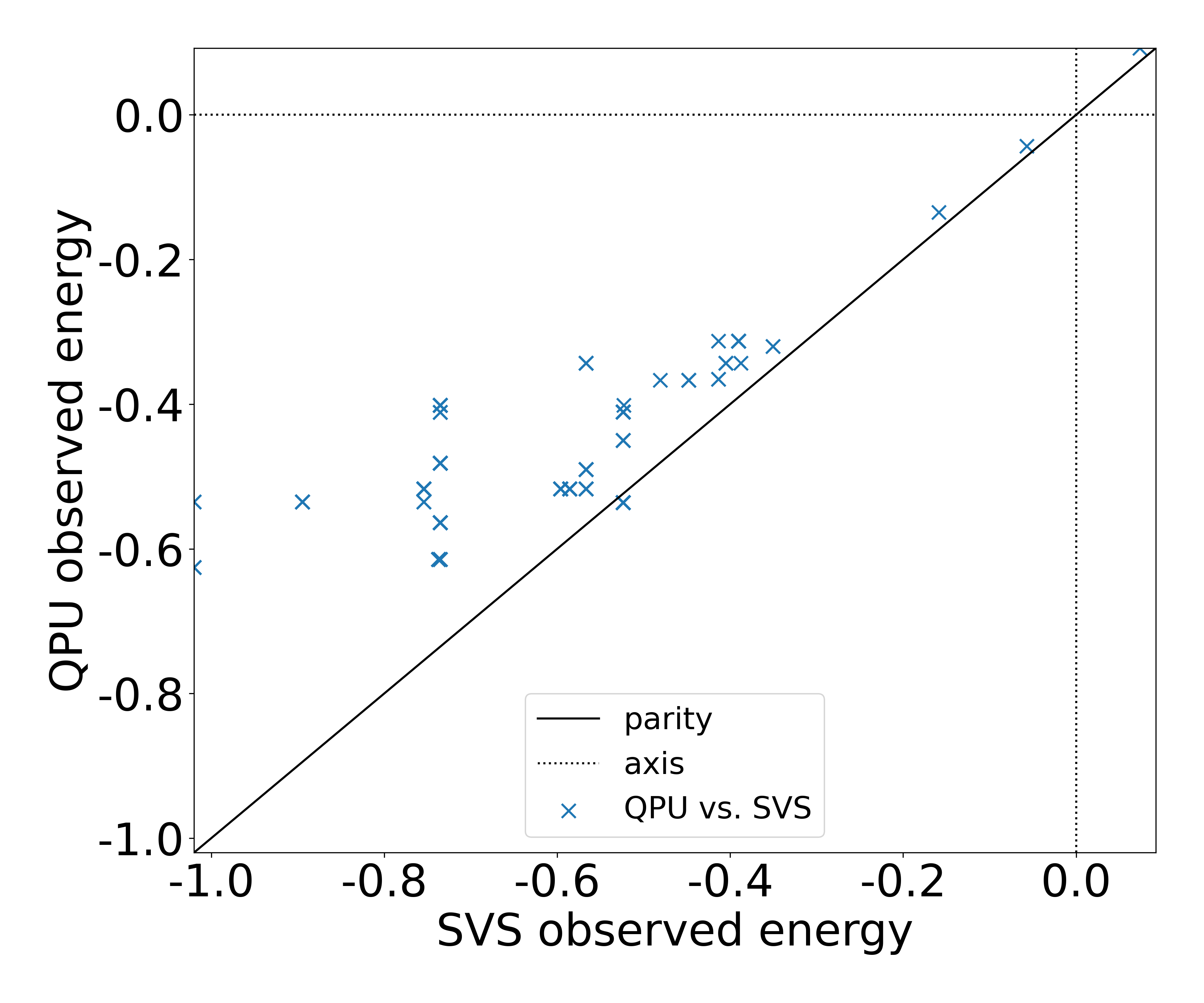}
    \caption{ (Left) Comparison of BOPT to Powell on quantum hardware simulations for $H_2$. Dotted lines are QPU optimization results on IBM's Torino, dashed lines are the SVS validation of the QPU trace trajectories, and solid lines denote the optimization on the noise-free SVS. (Right) Parity plot of the evaluated circuit parameters on IBM Torino versus the SVS.  }
    \label{fig:QPU_results}
\end{figure}

The results of the computation on quantum hardware are presented in Figure \ref{fig:QPU_results} (left).
A dotted line indicates the energies observed when optimizing on the QPU, and a dash-dotted line shows the energies from evaluating the QPU trace trajectories on an SVS.
The solid lines show the energies observed when optimizing on the SVS directly.
The QPU results show a bias similar to those observed on the fake backends, which is evident by similar changes in the trajectories of both option methods.
Powell and BOPT achieve similar performance on the QPU (when validated on the SVS) as they do on the SVS, suggesting that the variance of the noise does not significantly affect convergence.
Powell's resilience to hardware noise suggests it would be a suitable local optimization strategy to deploy from the best solution found by BOPT.
A parity plot between the QPU simulations and SVS is presented in Figure \ref{fig:QPU_results}  (right).
Relative to the fake backends, the QPU noise exhibits higher variance and appears to have a nonlinear bias.     

\section{Conclusions and Future Work}
\label{sec:conclusion}

In this work, we present a shot-efficient Bayesian optimization framework to address state preparation in the variational quantum eigensolver, referred to as Bayesian optimization with priors on topology (BOPT).
The framework uses a sparse Gaussian process model and low-shot circuit observations to construct a topological prior, where a residual is learned while performing Bayesian Optimization.
Statistical optimization results generated from state vector simulators (SVS), SVS with data-driven noise models, and a physical quantum computer show that this strategy significantly improves the standard Bayesian optimization algorithm and substantially outperforms the solver Powell, particularly when using periodic kernels in the Gaussian process models. 

Quantum computers are still too noisy for accurate ground state estimation; however, quantum computers may provide a computationally efficient means for state preparation using algorithms like the one presented here, which can be validated by other means, such as classical simulation.
Future work to be considered are noise mitigation strategies.
Several strategies are available, typically provided through the quantum computing interface, to mitigate hardware noise.
These methods typically incur a significant additional computational cost and may thus be reserved for validating prepared trial states, eliminating the need for any classical simulation.

Practical near-term quantum computing for chemistry and material science still requires further algorithmic developments.
Future work investigating modifications that allow efficient Bayesian optimization in high-dimensional parameter spaces is crucial to solving larger molecules of interest and simulating systems with more complex ans\"atze.
Furthermore, including geometry relaxation while solving VQE may increase the dimension of the optimization variables, depending on how the problem is formulated.
Local optimization methods deployed from the best solutions found with BOPT may make it possible to find solutions within chemical accuracy.
Although these modifications may make practical problems tractable on quantum computers, extensive benchmarking against classical methods will be needed to quantify the value of quantum computation of molecular ground state energies.

\section{Acknowledgements}

F.S. acknowledges funding support from the NSF Graduate Research Fellowship. 
M.T.R., N.E.B., and M.M.L. acknowledge the financial support of the European Union (Grant no. DCIPANAF/2020/420-028) through the African Research Initiative for Scientific Excellence (ARISE) pilot program.
J.A.P. acknowledges funding support from NSF Grant 2237616. 
D.B.N. is supported by the startup grant of the Davidson School of Chemical Engineering at Purdue University.
N.M.T. acknowledges support from the U.S. Department of Energy, Office of Science,
National Quantum Information Science Research Centers, Co-Design Center for Quantum Advantage under
Contract No. DE-SC0012704 (C2QA).

The authors are grateful for support from NASA Ames Research Center.
This research used resources of the National Energy Research
Scientific Computing Center, a DOE Office of Science User Facility supported by the Office of Science of the U.S. Department of Energy under Contract No. DE-AC02-05CH11231 using NERSC award ASCR-ERCAP0024469.
We acknowledge the use of IBM Quantum services through the Quantum Collaborative for this work.
The views expressed are those of the authors and do not reflect the official policy or position of IBM or the IBM Quantum team.
The authors thank the National Academies of Science, Engineering, and Medicine for their support through the Arab-American Frontiers Fellowship that enabled M.T.R.'s visit to D.B.N.

\bibliography{References}

\begin{thebibliography}{100}

\bibitem{ciapprox-2}
Norm~M. Tubman, C.~Daniel Freeman, Daniel~S. Levine, Diptarka Hait, Martin Head-Gordon, and K.~Birgitta Whaley.
\newblock {Modern Approaches to Exact Diagonalization and Selected Configuration Interaction with the Adaptive Sampling CI Method}.
\newblock {\em Journal of Chemical Theory and Computation}, 16(4):2139--2159, 2020.

\bibitem{ciapprox-book}
Attila Szabo and Neil~S Ostlund.
\newblock {\em {Modern quantum chemistry: introduction to advanced electronic structure theory}}.
\newblock Courier Corporation, 2012.

\bibitem{Faber17}
Felix~A. Faber, Luke Hutchison, Bing Huang, Justin Gilmer, Samuel~S. Schoenholz, George~E. Dahl, Oriol Vinyals, Steven Kearnes, Patrick~F. Riley, and O.~Anatole von Lilienfeld.
\newblock {Prediction Errors of Molecular Machine Learning Models Lower than Hybrid DFT Error}.
\newblock {\em Journal of Chemical Theory and Computation}, 13(11):5255--5264, 2017.

\bibitem{Li2024}
Zhengyuan Li, Peng Wang, Xiang Lyu, Vamsi Krishna~Reddy Kondapalli, Shuting Xiang, Juan~D. Jimenez, Lu~Ma, Takeshi Ito, Tianyu Zhang, Jithu Raj, Yanbo Fang, Yaocai Bai, Jianlin Li, Alexey Serov, Vesselin Shanov, Anatoly~I. Frenkel, Sanjaya~D. Senanayake, Shize Yang, Thomas~P. Senftle, and Jingjie Wu.
\newblock {Directing CO2 electroreduction pathways for selective C2 product formation using single-site doped copper catalysts}.
\newblock {\em Nature Chemical Engineering}, 1(2):159--169, February 2024.

\bibitem{Dietschreit}
Johannes C.~B. Dietschreit, Dennis~J. Diestler, and Rafael Gómez-Bombarelli.
\newblock {Entropy and Energy Profiles of Chemical Reactions}.
\newblock {\em Journal of Chemical Theory and Computation}, 19(16):5369--5379, 2023.
\newblock PMID: 37535443.

\bibitem{Nodozi}
Iman Nodozi, Charlie Yan, Mira Khare, Abhishek Halder, and Ali Mesbah.
\newblock {Neural Schrödinger Bridge With Sinkhorn Losses: Application to Data-Driven Minimum Effort Control of Colloidal Self-Assembly}.
\newblock {\em IEEE Transactions on Control Systems Technology}, 32(3):960--973, 2024.

\bibitem{Findeisen}
Rolf Findeisen, Martha~A. Grover, Christian Wagner, Michael Maiworm, Ruslan Temirov, F.~Stefan Tautz, Murti~V. Salapaka, Srinivasa Salapaka, Richard~D. Braatz, and S.O.~Reza Moheimani.
\newblock {Control on a molecular scale: A perspective}.
\newblock In {\em 2016 American Control Conference (ACC)}, pages 3069--3082, 2016.

\bibitem{BOONE2016221}
K.~Boone, F.~Abedin, M.R. Anwar, and K.V. Camarda.
\newblock {Chapter 8 - Molecular Design in the Pharmaceutical Industries}.
\newblock In Mariano Martín, Mario~R. Eden, and Nishanth~G. Chemmangattuvalappil, editors, {\em Tools For Chemical Product Design}, volume~39 of {\em Computer Aided Chemical Engineering}, pages 221--238. Elsevier, 2016.

\bibitem{Arora}
Akash Arora, David~C. Morse, Frank~S. Bates, and Kevin~D. Dorfman.
\newblock {Accelerating self-consistent field theory of block polymers in a variable unit cell}.
\newblock {\em The Journal of Chemical Physics}, 146(24):244902, 06 2017.

\bibitem{Chen2024}
Pengyu Chen and Kevin~D. Dorfman.
\newblock {A soft crystalline packing with no metallic analogue}.
\newblock {\em Nature Materials}, 23(4):455--456, April 2024.

\bibitem{Bui2024}
Justin~C. Bui, Eric~W. Lees, Daniela~H. Marin, T.~Nathan Stovall, Lihaokun Chen, Ahmet Kusoglu, Adam~C. Nielander, Thomas~F. Jaramillo, Shannon~W. Boettcher, Alexis~T. Bell, and Adam~Z. Weber.
\newblock {Multi-scale physics of bipolar membranes in electrochemical processes}.
\newblock {\em Nature Chemical Engineering}, 1(1):45--60, January 2024.

\bibitem{Bombarelli2016}
Rafael G{\'o}mez-Bombarelli, Jorge Aguilera-Iparraguirre, Timothy~D. Hirzel, Dong-Gwang Ha, Markus Einzinger, Tony Wu, Marc~A. Baldo, and Al{\'a}n Aspuru-Guzik.
\newblock {Turbocharged molecular discovery of OLED emitters: from high-throughput quantum simulation to highly efficient TADF devices}.
\newblock In Franky So, Chihaya Adachi, and Jang-Joo Kim, editors, {\em Organic Light Emitting Materials and Devices XX}, volume 9941, page 99410A. International Society for Optics and Photonics, SPIE, 2016.

\bibitem{BEFORT20221249}
Bridgette~J. Befort, Ryan~S. DeFever, Edward~J. Maginn, and Alexander~W. Dowling.
\newblock {Machine Learning-Enabled Optimization of Force Fields for Hydrofluorocarbons}.
\newblock In Yoshiyuki Yamashita and Manabu Kano, editors, {\em 14th International Symposium on Process Systems Engineering}, volume~49 of {\em Computer Aided Chemical Engineering}, pages 1249--1254. Elsevier, 2022.

\bibitem{ALEXEEV2024666}
Yuri Alexeev, Maximilian Amsler, Marco~Antonio Barroca, Sanzio Bassini, Torey Battelle, Daan Camps, David Casanova, Young~Jay Choi, Frederic~T. Chong, Charles Chung, Christopher Codella, Antonio~D. Córcoles, James Cruise, Alberto {Di Meglio}, Ivan Duran, Thomas Eckl, Sophia Economou, Stephan Eidenbenz, Bruce Elmegreen, Clyde Fare, Ismael Faro, Cristina~Sanz Fernández, Rodrigo Neumann~Barros Ferreira, Keisuke Fuji, Bryce Fuller, Laura Gagliardi, Giulia Galli, Jennifer~R. Glick, Isacco Gobbi, Pranav Gokhale, Salvador {de la Puente Gonzalez}, Johannes Greiner, Bill Gropp, Michele Grossi, Emanuel Gull, Burns Healy, Matthew~R. Hermes, Benchen Huang, Travis~S. Humble, Nobuyasu Ito, Artur~F. Izmaylov, Ali Javadi-Abhari, Douglas Jennewein, Shantenu Jha, Liang Jiang, Barbara Jones, Wibe~Albert {de Jong}, Petar Jurcevic, William Kirby, Stefan Kister, Masahiro Kitagawa, Joel Klassen, Katherine Klymko, Kwangwon Koh, Masaaki Kondo, Doga~Murat Kurkcuoglu, Krzysztof Kurowski, Teodoro Laino, Ryan Landfield, Matt Leininger,
  Vicente Leyton-Ortega, Ang Li, Meifeng Lin, Junyu Liu, Nicolas Lorente, Andre Luckow, Simon Martiel, Francisco Martin-Fernandez, Margaret Martonosi, Claire Marvinney, Arcesio~Castaneda Medina, Dirk Merten, Antonio Mezzacapo, Kristel Michielsen, Abhishek Mitra, Tushar Mittal, Kyungsun Moon, Joel Moore, Sarah Mostame, Mario Motta, Young-Hye Na, Yunseong Nam, Prineha Narang, Yu~ya~Ohnishi, Daniele Ottaviani, Matthew Otten, Scott Pakin, Vincent~R. Pascuzzi, Edwin Pednault, Tomasz Piontek, Jed Pitera, Patrick Rall, Gokul~Subramanian Ravi, Niall Robertson, Matteo~A.C. Rossi, Piotr Rydlichowski, Hoon Ryu, Georgy Samsonidze, Mitsuhisa Sato, Nishant Saurabh, Vidushi Sharma, Kunal Sharma, Soyoung Shin, George Slessman, Mathias Steiner, Iskandar Sitdikov, In-Saeng Suh, Eric~D. Switzer, Wei Tang, Joel Thompson, Synge Todo, Minh~C. Tran, Dimitar Trenev, Christian Trott, Huan-Hsin Tseng, Norm~M. Tubman, Esin Tureci, David~García Valiñas, Sofia Vallecorsa, Christopher Wever, Konrad Wojciechowski, Xiaodi Wu, Shinjae Yoo,
  Nobuyuki Yoshioka, Victor~Wen zhe Yu, Seiji Yunoki, Sergiy Zhuk, and Dmitry Zubarev.
\newblock Quantum-centric supercomputing for materials science: A perspective on challenges and future directions.
\newblock {\em Future Generation Computer Systems}, 160:666--710, 2024.

\bibitem{Bassman2021}
Lindsay~Bassman Oftelie, Miroslav Urbanek, Mekena Metcalf, Jonathan Carter, Alexander~F Kemper, and Wibe~A de~Jong.
\newblock Simulating quantum materials with digital quantum computers.
\newblock {\em Quantum Science and Technology}, 6(4):043002, sep 2021.

\bibitem{Kwon}
Soonho Kwon, Kelsey~A. Stoerzinger, Reshma Rao, Liang Qiao, William A.~III Goddard, and Yang Shao-Horn.
\newblock {Facet-Dependent Oxygen Evolution Reaction Activity of IrO2 from Quantum Mechanics and Experiments}.
\newblock {\em Journal of the American Chemical Society}, 146(17):11719--11725, 2024.
\newblock PMID: 38636103.

\bibitem{Goldsmith}
Bryan~R. Goldsmith, Jacques Esterhuizen, Jin-Xun Liu, Christopher~J. Bartel, and Christopher Sutton.
\newblock {Machine learning for heterogeneous catalyst design and discovery}.
\newblock {\em AIChE Journal}, 64(7):2311--2323, 2018.

\bibitem{NIKBIN201335}
Nima Nikbin, Phuong~T. Do, Stavros Caratzoulas, Raul~F. Lobo, Paul~J. Dauenhauer, and Dionisios~G. Vlachos.
\newblock {A DFT study of the acid-catalyzed conversion of 2,5-dimethylfuran and ethylene to p-xylene}.
\newblock {\em Journal of Catalysis}, 297:35--43, 2013.

\bibitem{Adjiman21}
Claire~S. Adjiman, Nikolaos~V. Sahinidis, Dionisios~G. Vlachos, Bhavik Bakshi, Christos~T. Maravelias, and Christos Georgakis.
\newblock {Process Systems Engineering Perspective on the Design of Materials and Molecules}.
\newblock {\em Industrial \& Engineering Chemistry Research}, 60(14):5194--5206, 2021.

\bibitem{Derezinski}
Jan Derezinski.
\newblock {Asymptotic Completeness of Long-Range N-Body Quantum Systems}.
\newblock {\em Annals of Mathematics}, 138(2):427--476, 1993.

\bibitem{Guzik2005}
Aspuru-Guzik Alan, Anthony~D. Dutoi, Peter~J. Love, and Martin Head-Gordon.
\newblock {Simulated Quantum Computation of Molecular Energies}.
\newblock {\em Science}, September 2005.

\bibitem{eriksen2020ground}
Janus~J Eriksen, Tyler~A Anderson, J~Emiliano Deustua, Khaldoon Ghanem, Diptarka Hait, Mark~R Hoffmann, Seunghoon Lee, Daniel~S Levine, Ilias Magoulas, Jun Shen, et~al.
\newblock {The ground state electronic energy of benzene}.
\newblock {\em The journal of physical chemistry letters}, 11(20):8922--8929, 2020.

\bibitem{Kim_2018}
Jeongnim Kim, Andrew~D Baczewski, Todd~D Beaudet, Anouar Benali, M~Chandler Bennett, Mark~A Berrill, Nick~S Blunt, Edgar Josué~Landinez Borda, Michele Casula, David~M Ceperley, Simone Chiesa, Bryan~K Clark, Raymond~C Clay, Kris~T Delaney, Mark Dewing, Kenneth~P Esler, Hongxia Hao, Olle Heinonen, Paul R~C Kent, Jaron~T Krogel, Ilkka Kylänpää, Ying~Wai Li, M~Graham Lopez, Ye~Luo, Fionn~D Malone, Richard~M Martin, Amrita Mathuriya, Jeremy McMinis, Cody~A Melton, Lubos Mitas, Miguel~A Morales, Eric Neuscamman, William~D Parker, Sergio D~Pineda Flores, Nichols~A Romero, Brenda~M Rubenstein, Jacqueline A~R Shea, Hyeondeok Shin, Luke Shulenburger, Andreas~F Tillack, Joshua~P Townsend, Norm~M Tubman, Brett Van~Der Goetz, Jordan~E Vincent, D~ChangMo Yang, Yubo Yang, Shuai Zhang, and Luning Zhao.
\newblock {QMCPACK: an open source ab initio quantum Monte Carlo package for the electronic structure of atoms, molecules and solids}.
\newblock {\em Journal of Physics: Condensed Matter}, 30(19):195901, April 2018.

\bibitem{ciapprox-1}
Norm~M. Tubman, Joonho Lee, Tyler~Y. Takeshita, Martin Head-Gordon, and K.~Birgitta Whaley.
\newblock {A deterministic alternative to the full configuration interaction quantum Monte Carlo method}.
\newblock {\em The Journal of Chemical Physics}, 145(4):044112, 2016.

\bibitem{Bogojeski2020}
Mihail Bogojeski, Leslie Vogt-Maranto, Mark~E. Tuckerman, Klaus-Robert M{\"u}ller, and Kieron Burke.
\newblock {Quantum chemical accuracy from density functional approximations via machine learning}.
\newblock {\em Nature Communications}, 11(1):5223, October 2020.

\bibitem{Shi23}
Yi~Shi, Yuming Shi, and Adam Wasserman.
\newblock {Strong electron correlation from partition density functional theory}.
\newblock {\em The Journal of Chemical Physics}, 159(22):224108, December 2023.

\bibitem{TILLY20221}
Jules Tilly, Hongxiang Chen, Shuxiang Cao, Dario Picozzi, Kanav Setia, Ying Li, Edward Grant, Leonard Wossnig, Ivan Rungger, George~H. Booth, and Jonathan Tennyson.
\newblock {The Variational Quantum Eigensolver: A review of methods and best practices}.
\newblock {\em Physics Reports}, 986:1--128, 2022.
\newblock The Variational Quantum Eigensolver: a review of methods and best practices.

\bibitem{Fauseweh2024}
Benedikt Fauseweh.
\newblock {Quantum many-body simulations on digital quantum computers: State-of-the-art and future challenges}.
\newblock {\em Nature Communications}, 15(1):2123, March 2024.

\bibitem{gustafson2024surrogate}
Erik~J Gustafson, Juha Tiihonen, Diana Chamaki, Farshud Sorourifar, J~Wayne Mullinax, Andy~CY Li, Filip~B Maciejewski, Nicolas~PD Sawaya, Jaron~T Krogel, David~E {Bernal Neira}, and Norman Tubman.
\newblock {Surrogate optimization of variational quantum circuits}, 2024.

\bibitem{Grimsley_2023}
Harper~R. Grimsley, George~S. Barron, Edwin Barnes, Sophia~E. Economou, and Nicholas~J. Mayhall.
\newblock {Adaptive, problem-tailored variational quantum eigensolver mitigates rough parameter landscapes and barren plateaus}.
\newblock {\em npj Quantum Information}, 9(1), March 2023.

\bibitem{Anschuetz_2022}
Eric~R. Anschuetz and Bobak~T. Kiani.
\newblock {Quantum variational algorithms are swamped with traps}.
\newblock {\em Nature Communications}, 13(1), December 2022.

\bibitem{Kandala2017}
Abhinav Kandala, Antonio Mezzacapo, Kristan Temme, Maika Takita, Markus Brink, Jerry~M. Chow, and Jay~M. Gambetta.
\newblock {Hardware-efficient variational quantum eigensolver for small molecules and quantum magnets}.
\newblock {\em Nature}, 549(7671):242--246, September 2017.

\bibitem{PRXQuantum.3.020323}
Katherine Klymko, Carlos Mejuto-Zaera, Stephen~J. Cotton, Filip Wudarski, Miroslav Urbanek, Diptarka Hait, Martin Head-Gordon, K.~Birgitta Whaley, Jonathan Moussa, Nathan Wiebe, Wibe~A. de~Jong, and Norm~M. Tubman.
\newblock Real-time evolution for ultracompact hamiltonian eigenstates on quantum hardware.
\newblock {\em PRX Quantum}, 3:020323, May 2022.

\bibitem{PhysRevResearch.5.033071}
Andy C.~Y. Li, M.~Sohaib Alam, Thomas Iadecola, Ammar Jahin, Joshua Job, Doga~Murat Kurkcuoglu, Richard Li, Peter~P. Orth, A.~Bar\ifmmode \imath \else \i \fi{}\ifmmode \mbox{\c{s}}\else~\c{s}\fi{} \"Ozg\"uler, Gabriel~N. Perdue, and Norm~M. Tubman.
\newblock Benchmarking variational quantum eigensolvers for the square-octagon-lattice kitaev model.
\newblock {\em Phys. Rev. Res.}, 5:033071, Aug 2023.

\bibitem{darbha2024false}
Siva Darbha, Milan Kornjača, Fangli Liu, Jan Balewski, Mark~R. Hirsbrunner, Pedro Lopes, Sheng-Tao Wang, Roel~Van Beeumen, Daan Camps, and Katherine Klymko.
\newblock False vacuum decay and nucleation dynamics in neutral atom systems, 2024.

\bibitem{darbha2024long}
Siva Darbha, Milan Kornjača, Fangli Liu, Jan Balewski, Mark~R. Hirsbrunner, Pedro Lopes, Sheng-Tao Wang, Roel~Van Beeumen, Katherine Klymko, and Daan Camps.
\newblock Long-lived oscillations of false and true vacuum states in neutral atom systems, 2024.

\bibitem{2024arXiv24beyond}
Javier {Robledo-Moreno}, Mario {Motta}, Holger {Haas}, Ali {Javadi-Abhari}, Petar {Jurcevic}, William {Kirby}, Simon {Martiel}, Kunal {Sharma}, Sandeep {Sharma}, Tomonori {Shirakawa}, Iskandar {Sitdikov}, Rong-Yang {Sun}, Kevin~J. {Sung}, Maika {Takita}, Minh~C. {Tran}, Seiji {Yunoki}, and Antonio {Mezzacapo}.
\newblock {Chemistry Beyond Exact Solutions on a Quantum-Centric Supercomputer}.
\newblock {\em arXiv e-prints}, page arXiv:2405.05068, May 2024.

\bibitem{arute2020observ}
Frank Arute, Kunal Arya, Ryan Babbush, Dave Bacon, Joseph~C. Bardin, Rami Barends, Andreas Bengtsson, Sergio Boixo, Michael Broughton, Bob~B. Buckley, David~A. Buell, Brian Burkett, Nicholas Bushnell, Yu~Chen, Zijun Chen, Yu-An Chen, Ben Chiaro, Roberto Collins, Stephen~J. Cotton, William Courtney, Sean Demura, Alan Derk, Andrew Dunsworth, Daniel Eppens, Thomas Eckl, Catherine Erickson, Edward Farhi, Austin Fowler, Brooks Foxen, Craig Gidney, Marissa Giustina, Rob Graff, Jonathan~A. Gross, Steve Habegger, Matthew~P. Harrigan, Alan Ho, Sabrina Hong, Trent Huang, William Huggins, Lev~B. Ioffe, Sergei~V. Isakov, Evan Jeffrey, Zhang Jiang, Cody Jones, Dvir Kafri, Kostyantyn Kechedzhi, Julian Kelly, Seon Kim, Paul~V. Klimov, Alexander~N. Korotkov, Fedor Kostritsa, David Landhuis, Pavel Laptev, Mike Lindmark, Erik Lucero, Michael Marthaler, Orion Martin, John~M. Martinis, Anika Marusczyk, Sam McArdle, Jarrod~R. McClean, Trevor McCourt, Matt McEwen, Anthony Megrant, Carlos Mejuto-Zaera, Xiao Mi, Masoud Mohseni,
  Wojciech Mruczkiewicz, Josh Mutus, Ofer Naaman, Matthew Neeley, Charles Neill, Hartmut Neven, Michael Newman, Murphy~Yuezhen Niu, Thomas~E. O'Brien, Eric Ostby, Bálint Pató, Andre Petukhov, Harald Putterman, Chris Quintana, Jan-Michael Reiner, Pedram Roushan, Nicholas~C. Rubin, Daniel Sank, Kevin~J. Satzinger, Vadim Smelyanskiy, Doug Strain, Kevin~J. Sung, Peter Schmitteckert, Marco Szalay, Norm~M. Tubman, Amit Vainsencher, Theodore White, Nicolas Vogt, Z.~Jamie Yao, Ping Yeh, Adam Zalcman, and Sebastian Zanker.
\newblock Observation of separated dynamics of charge and spin in the fermi-hubbard model, 2020.

\bibitem{ollitrault2024estimation}
Pauline~J Ollitrault, Matthias Loipersberger, Robert~M Parrish, Alexander Erhard, Christine Maier, Christian Sommer, Juris Ulmanis, Thomas Monz, Christian Gogolin, Christofer~S Tautermann, et~al.
\newblock Estimation of electrostatic interaction energies on a trapped-ion quantum computer.
\newblock {\em ACS Central Science}, 10(4):882--889, 2024.

\bibitem{anastasiou2022tetrisadaptvqe}
Panagiotis~G. Anastasiou, Yanzhu Chen, Nicholas~J. Mayhall, Edwin Barnes, and Sophia~E. Economou.
\newblock {TETRIS-ADAPT-VQE: An adaptive algorithm that yields shallower, denser circuit ans\"atze}.
\newblock {\em Physical Review Research}, 6(1):013254, 2024.

\bibitem{2016NJPh...18b3023M}
Jarrod~R. {McClean}, Jonathan {Romero}, Ryan {Babbush}, and Al{\'a}n {Aspuru-Guzik}.
\newblock {The theory of variational hybrid quantum-classical algorithms}.
\newblock {\em New Journal of Physics}, 18(2):023023, February 2016.

\bibitem{bittel2022fast}
Lennart Bittel, Jens Watty, and Martin Kliesch.
\newblock {Fast gradient estimation for variational quantum algorithms}.
\newblock {\em arXiv preprint arXiv:2210.06484}, 2022.

\bibitem{PhysRevA.98.032309}
K.~Mitarai, M.~Negoro, M.~Kitagawa, and K.~Fujii.
\newblock {Quantum circuit learning}.
\newblock {\em Phys. Rev. A}, 98:032309, September 2018.

\bibitem{Harrow2021}
Aram~W. Harrow and John~C. Napp.
\newblock {Low-Depth Gradient Measurements Can Improve Convergence in Variational Hybrid Quantum-Classical Algorithms}.
\newblock {\em Phys. Rev. Lett.}, 126:140502, April 2021.

\bibitem{jones2020efficient}
Tyson Jones and Julien Gacon.
\newblock {Efficient calculation of gradients in classical simulations of variational quantum algorithms}.
\newblock {\em arXiv preprint arXiv:2009.02823}, 2020.

\bibitem{Berg2022}
Erik Berglund, Sarit Khirirat, and Xiaoyu Wang.
\newblock {Zeroth-Order Randomized Subspace Newton Methods}.
\newblock In {\em ICASSP 2022 - 2022 IEEE International Conference on Acoustics, Speech and Signal Processing (ICASSP)}, pages 6002--6006, 2022.

\bibitem{zhew2020}
Zhewei {Yao}, Amir {Gholami}, Sheng {Shen}, Mustafa {Mustafa}, Kurt {Keutzer}, and Michael~W. {Mahoney}.
\newblock {ADAHESSIAN: An Adaptive Second Order Optimizer for Machine Learning}.
\newblock {\em arXiv e-prints}, page arXiv:2006.00719, June 2020.

\bibitem{PhysRevA.107.032415}
Ryan Shaffer, Lucas Kocia, and Mohan Sarovar.
\newblock {Surrogate-based optimization for variational quantum algorithms}.
\newblock {\em Phys. Rev. A}, 107:032415, March 2023.

\bibitem{hirsbrunner2024diag}
Mark~R. Hirsbrunner, J.~Wayne Mullinax, Yizhi Shen, David~B. Williams-Young, Katherine Klymko, Roel~Van Beeumen, and Norm~M. Tubman.
\newblock Diagnosing local minima and accelerating convergence of variational quantum eigensolvers with quantum subspace techniques, 2024.

\bibitem{leimkuhler2024}
Oskar {Leimkuhler} and K.~Birgitta {Whaley}.
\newblock {A quantum eigenvalue solver based on tensor networks}.
\newblock {\em arXiv e-prints}, page arXiv:2404.10223, April 2024.

\bibitem{ollitrault2024enhancing}
Pauline~J Ollitrault, Cristian~L Cortes, Jerome~F Gonthier, Robert~M Parrish, Dario Rocca, Gian-Luca Anselmetti, Matthias Degroote, Nikolaj Moll, Raffaele Santagati, and Michael Streif.
\newblock Enhancing initial state overlap through orbital optimization for faster molecular electronic ground-state energy estimation.
\newblock {\em arXiv preprint arXiv:2404.08565}, 2024.

\bibitem{2024arXivruslan}
Tianyi {Hao}, Zichang {He}, Ruslan {Shaydulin}, Marco {Pistoia}, and Swamit {Tannu}.
\newblock {Variational Quantum Algorithm Landscape Reconstruction by Low-Rank Tensor Completion}.
\newblock {\em arXiv e-prints}, page arXiv:2405.10941, May 2024.

\bibitem{Fin_gar_2024}
Jernej~Rudi Finžgar, Martin J.~A. Schuetz, J.~Kyle Brubaker, Hidetoshi Nishimori, and Helmut~G. Katzgraber.
\newblock Designing quantum annealing schedules using bayesian optimization.
\newblock {\em Physical Review Research}, 6(2), April 2024.

\bibitem{Cheng_2024}
Lixue Cheng, Yu-Qin Chen, Shi-Xin Zhang, and Shengyu Zhang.
\newblock Quantum approximate optimization via learning-based adaptive optimization.
\newblock {\em Communications Physics}, 7(1), March 2024.

\bibitem{otterbach2017unsupervised}
J.~S. Otterbach, R.~Manenti, N.~Alidoust, A.~Bestwick, M.~Block, B.~Bloom, S.~Caldwell, N.~Didier, E.~Schuyler Fried, S.~Hong, P.~Karalekas, C.~B. Osborn, A.~Papageorge, E.~C. Peterson, G.~Prawiroatmodjo, N.~Rubin, Colm~A. Ryan, D.~Scarabelli, M.~Scheer, E.~A. Sete, P.~Sivarajah, Robert~S. Smith, A.~Staley, N.~Tezak, W.~J. Zeng, A.~Hudson, Blake~R. Johnson, M.~Reagor, M.~P. da~Silva, and C.~Rigetti.
\newblock {Unsupervised Machine Learning on a Hybrid Quantum Computer}.
\newblock {\em arXiv preprint arXiv:1712.05771}, 2017.

\bibitem{tibaldi23}
Simone Tibaldi, Davide Vodola, Edoardo Tignone, and Elisa Ercolessi.
\newblock {Bayesian Optimization for QAOA}.
\newblock {\em IEEE Transactions on Quantum Engineering}, 2023.

\bibitem{Ciavarella22}
Anthony~N. Ciavarella and Ivan~A. Chernyshev.
\newblock {Preparation of the SU(3) lattice Yang-Mills vacuum with variational quantum methods}.
\newblock {\em Phys. Rev. D}, 105:074504, April 2022.

\bibitem{iannelli2021noisy}
Giovanni Iannelli and Karl Jansen.
\newblock Noisy bayesian optimization for variational quantum eigensolvers.
\newblock {\em arXiv preprint arxiv:2112.00426}, 2021.

\bibitem{Eriksson19}
David Eriksson, Michael Pearce, Jacob Gardner, Ryan~D Turner, and Matthias Poloczek.
\newblock {Scalable Global Optimization via Local Bayesian Optimization}.
\newblock In {\em Advances in Neural Information Processing Systems}, volume~32. Curran Associates, Inc., 2019.

\bibitem{Rasmussen2006}
C.~E. Rasmussen and C.~K.~I. Williams.
\newblock {\em {Gaussian Processes for Machine Learning.}}
\newblock MIT Press, 2006.

\bibitem{nei}
Benjamin Letham, Brian Karrer, Guilherme Ottoni, and Eytan Bakshy.
\newblock {Constrained Bayesian Optimization with Noisy Experiments}.
\newblock {\em Bayesian Analysis}, 14(2):495 -- 519, 2019.

\bibitem{jones1998efficient}
Donald~R Jones, Matthias Schonlau, and William~J Welch.
\newblock {Efficient global optimization of expensive black-box functions}.
\newblock {\em Journal of Global Optimization}, 13:455--492, 1998.

\bibitem{balandat2020botorch}
Maximilian Balandat, Brian Karrer, Daniel Jiang, Samuel Daulton, Ben Letham, Andrew~G Wilson, and Eytan Bakshy.
\newblock {BoTorch: A Framework for Efficient Monte-Carlo Bayesian Optimization}.
\newblock {\em Advances in Neural Information Processing Systems}, 33:21524--21538, 2020.

\bibitem{frazier18}
P.~I. Frazier.
\newblock {A Tutorial on Bayesian Optimization}.
\newblock {\em arXiv preprint arXiv:1807.02811}, 2018.

\bibitem{Muller22}
J.~Muller, W.~Lavrijsen, C.~Iancu, and W.~de~Jong.
\newblock {Accelerating Noisy VQE Optimization with Gaussian Processes}.
\newblock In {\em 2022 IEEE International Conference on Quantum Computing and Engineering (QCE)}, pages 215--225. IEEE Computer Society, September 2022.

\bibitem{Tamiya2022}
Shiro Tamiya and Hayata Yamasaki.
\newblock {Stochastic gradient line Bayesian optimization for efficient noise-robust optimization of parameterized quantum circuits}.
\newblock {\em npj Quantum Information}, 8(1):90, July 2022.

\bibitem{Shaffer2023}
Ryan Shaffer, Lucas Kocia, and Mohan Sarovar.
\newblock {Surrogate-based optimization for variational quantum algorithms}.
\newblock {\em Phys. Rev. A}, 107:032415, March 2023.

\bibitem{Cheng2024}
Lixue Cheng, Yu-Qin Chen, Shi-Xin Zhang, and Shengyu Zhang.
\newblock {Quantum approximate optimization via learning-based adaptive optimization}.
\newblock {\em Communications Physics}, 7(1):83, March 2024.

\bibitem{Self2021}
Chris~N. Self, Kiran~E. Khosla, Alistair W.~R. Smith, Fr{\'e}d{\'e}ric Sauvage, Peter~D. Haynes, Johannes Knolle, Florian Mintert, and M.~S. Kim.
\newblock {Variational quantum algorithm with information sharing}.
\newblock {\em npj Quantum Information}, 7(1):116, July 2021.

\bibitem{rohrs2024bayesian}
Milena R{\"o}hrs, Alexey Bochkarev, and Arcesio~C Medina.
\newblock {Bayesian optimisation with improved information sharing for the variational quantum eigensolver}.
\newblock {\em arXiv preprint arXiv:2405.14353}, 2024.

\bibitem{Nicoli2023}
Kim Nicoli, Christopher~J. Anders, Lena Funcke, Tobias Hartung, Karl Jansen, Stefan K\"{u}hn, Klaus-Robert M\"{u}ller, Paolo Stornati, Pan Kessel, and Shinichi Nakajima.
\newblock {Physics-Informed Bayesian Optimization of Variational Quantum Circuits}.
\newblock In A.~Oh, T.~Naumann, A.~Globerson, K.~Saenko, M.~Hardt, and S.~Levine, editors, {\em Advances in Neural Information Processing Systems}, volume~36, pages 18341--18376. Curran Associates, Inc., 2023.

\bibitem{Nakanishi2020}
Ken~M. Nakanishi, Keisuke Fujii, and Synge Todo.
\newblock {Sequential minimal optimization for quantum-classical hybrid algorithms}.
\newblock {\em Phys. Rev. Res.}, 2:043158, October 2020.

\bibitem{escape}
F~Sorourifar, D~Chamaki, J~Paulson, N~Tubman, and D.E. {Bernal Neira}.
\newblock {Bayesian Optimization Priors for Efficient Variational Quantum Algorithms}.
\newblock In {\em Proceedings of the 34th European Symposium on Computer Aided Process Engineering / 15th International Symposium on Process Systems Engineering (ESCAPE34/PSE2024)}, 2024.
\newblock To be published.

\bibitem{Bernal22}
David~E. Bernal, Akshay Ajagekar, Stuart~M. Harwood, Spencer~T. Stober, Dimitar Trenev, and Fengqi You.
\newblock {Perspectives of quantum computing for chemical engineering}.
\newblock {\em AIChE Journal}, 68(6):e17651, 2022.

\bibitem{Castelvecchi2017}
Davide Castelvecchi.
\newblock {Quantum computers ready to leap out of the lab in 2017}.
\newblock {\em Nature}, 541(7635):9--10, January 2017.

\bibitem{Linke}
Norbert~M. Linke, Dmitri Maslov, Martin Roetteler, Shantanu Debnath, Caroline Figgatt, Kevin~A. Landsman, Kenneth Wright, and Christopher Monroe.
\newblock {Experimental comparison of two quantum computing architectures}.
\newblock {\em Proceedings of the National Academy of Sciences}, 114(13):3305--3310, 2017.

\bibitem{nielsen2010quantum}
Michael~A Nielsen and Isaac~L Chuang.
\newblock {\em {Quantum computation and quantum information}}.
\newblock Cambridge university press, 2010.

\bibitem{Dahlhauser}
Megan~L. Dahlhauser and Travis~S. Humble.
\newblock {Benchmarking characterization methods for noisy quantum circuits}.
\newblock {\em Phys. Rev. A}, 109:042620, April 2024.

\bibitem{Preskill2018quantumcomputingin}
John Preskill.
\newblock {Quantum Computing in the NISQ era and beyond}.
\newblock {\em Quantum}, 2:79, August 2018.

\bibitem{kim2023evidence}
Youngseok Kim, Andrew Eddins, Sajant Anand, Ken~Xuan Wei, Ewout Van Den~Berg, Sami Rosenblatt, Hasan Nayfeh, Yantao Wu, Michael Zaletel, Kristan Temme, et~al.
\newblock {Evidence for the utility of quantum computing before fault tolerance}.
\newblock {\em Nature}, 618(7965):500--505, 2023.

\bibitem{shor1995a}
Peter~W. Shor.
\newblock {Scheme for reducing decoherence in quantum computer memory}.
\newblock {\em Phys. Rev. A}, 52:R2493--R2496, October 1995.

\bibitem{Kandala2019}
Abhinav Kandala, Kristan Temme, Antonio~D. C{\'o}rcoles, Antonio Mezzacapo, Jerry~M. Chow, and Jay~M. Gambetta.
\newblock {Error mitigation extends the computational reach of a noisy quantum processor}.
\newblock {\em Nature}, 567(7749):491--495, March 2019.

\bibitem{Wu2021}
Anbang Wu, Gushu Li, Yuke Wang, Boyuan Feng, Yufei Ding, and Yuan Xie.
\newblock {Towards Efficient Ansatz Architecture for Variational Quantum Algorithms}.
\newblock {\em arXiv preprint arXiv:2111.13730}, 2021.

\bibitem{xiao2023physicsconstrained}
Xiaoxiao Xiao, Hewang Zhao, Jiajun Ren, Wei-Hai Fang, and Zhendong Li.
\newblock {Physics-Constrained Hardware-Efficient Ansatz on Quantum Computers that is Universal, Systematically Improvable, and Size-consistent}.
\newblock {\em Journal of Chemical Theory and Computation}, 2024.

\bibitem{peruzzo2014variational}
Alberto Peruzzo, Jarrod McClean, Peter Shadbolt, Man-Hong Yung, Xiao-Qi Zhou, Peter~J Love, Al{\'a}n Aspuru-Guzik, and Jeremy~L O’brien.
\newblock {A variational eigenvalue solver on a photonic quantum processor}.
\newblock {\em Nature communications}, 5(1):4213, 2014.

\bibitem{anastasiou2023really}
Panagiotis~G Anastasiou, Nicholas~J Mayhall, Edwin Barnes, and Sophia~E Economou.
\newblock {How to really measure operator gradients in ADAPT-VQE}.
\newblock {\em arXiv preprint arXiv:2306.03227}, 2023.

\bibitem{romero2018strategies}
Jonathan Romero, Ryan Babbush, Jarrod~R McClean, Cornelius Hempel, Peter~J Love, and Al{\'a}n Aspuru-Guzik.
\newblock {Strategies for quantum computing molecular energies using the unitary coupled cluster ansatz}.
\newblock {\em Quantum Science and Technology}, 4(1):014008, 2018.

\bibitem{Cao2019}
Yudong Cao, Jonathan Romero, Jonathan~P. Olson, Matthias Degroote, Peter~D. Johnson, M{\'{a} }ria Kieferov{\'{a}}, Ian~D. Kivlichan, Tim Menke, Borja Peropadre, Nicolas P.~D. Sawaya, Sukin Sim, Libor Veis, and Al{\'{a}}n Aspuru-Guzik.
\newblock {Quantum Chemistry in the Age of Quantum Computing}.
\newblock {\em Chemical Reviews}, 119(19):10856--10915, August 2019.

\bibitem{Anand_2022}
Abhinav Anand, Philipp Schleich, Sumner Alperin-Lea, Phillip W.~K. Jensen, Sukin Sim, Manuel Díaz-Tinoco, Jakob~S. Kottmann, Matthias Degroote, Artur~F. Izmaylov, and Alán Aspuru-Guzik.
\newblock {A quantum computing view on unitary coupled cluster theory}.
\newblock {\em Chemical Society Reviews}, 51(5):1659–1684, 2022.

\bibitem{wang2014theoretical}
Z.~Wang and N.~de~Freitas.
\newblock {Theoretical analysis of Bayesian optimisation with unknown Gaussian process hyperparameters}.
\newblock {\em arXiv preprint arXiv:1406.7758}, 2014.

\bibitem{2017arXiv170401127H}
Thomas {H{\"a}ner} and Damian~S. {Steiger}.
\newblock {0.5 Petabyte Simulation of a 45-Qubit Quantum Circuit}.
\newblock {\em arXiv e-prints}, page arXiv:1704.01127, April 2017.

\bibitem{mullinax2023large}
J~Wayne Mullinax and Norm~M Tubman.
\newblock {Large-scale sparse wavefunction circuit simulator for applications with the variational quantum eigensolver}.
\newblock {\em arXiv preprint arXiv:2301.05726}, 2023.

\bibitem{2023arXiv231012965K}
Abid {Khan}, Bryan~K. {Clark}, and Norm~M. {Tubman}.
\newblock {Pre-optimizing variational quantum eigensolvers with tensor networks}.
\newblock {\em arXiv e-prints}, page arXiv:2310.12965, October 2023.

\bibitem{lu2021bayesian}
Qiugang Lu, Leonardo~D Gonz{\'a}lez, Ranjeet Kumar, and Victor~M Zavala.
\newblock {Bayesian Optimization with Reference Models: A Case Study in MPC for HVAC Central Plants}.
\newblock {\em Computers \& Chemical Engineering}, 154:107491, 2021.

\bibitem{aws_price}
Amazon.
\newblock {Amazon Web Services, Amazon Braket Pricing}, 2024.

\bibitem{Azure_price}
Microsoft.
\newblock {Microsoft Build, Azure Quantum Pricing}, 2024.

\bibitem{titsias2009svgp}
Michalis Titsias.
\newblock {Variational Learning of Inducing Variables in Sparse Gaussian Processes}.
\newblock In David van Dyk and Max Welling, editors, {\em Proceedings of the Twelth International Conference on Artificial Intelligence and Statistics}, volume~5 of {\em Proceedings of Machine Learning Research}, pages 567--574. PMLR, 16--18 Apr 2009.

\bibitem{Hensman2013svgp}
James Hensman, Nicol{\'{o}} Fusi, and Neil~D. Lawrence.
\newblock {Gaussian Processes for Big Data}.
\newblock {\em CoRR}, abs/1309.6835, 2013.

\bibitem{Burt2020svgp}
David~R. Burt, Carl~Edward Rasmussen, and Mark van~der Wilk.
\newblock {Convergence of Sparse Variational Inference in Gaussian Processes Regression}.
\newblock {\em Journal of Machine Learning Research}, 21(131):1--63, 2020.

\bibitem{moss2023svgp}
Henry~B Moss, Sebastian~W Ober, and Victor Picheny.
\newblock {Inducing point allocation for sparse Gaussian processes in high-throughput Bayesian optimisation}.
\newblock In {\em International Conference on Artificial Intelligence and Statistics}, pages 5213--5230. PMLR, 2023.

\bibitem{NaturalGradients}
Shun-ichi Amari.
\newblock {Natural Gradient Works Efficiently in Learning}.
\newblock {\em Neural Computation}, 10(2):251--276, 1998.

\bibitem{powell1964efficient}
Michael~JD Powell.
\newblock {An efficient method for finding the minimum of a function of several variables without calculating derivatives}.
\newblock {\em The computer journal}, 7(2):155--162, 1964.

\bibitem{Singh23}
Harshdeep Singh, Sonjoy Majumder, and Sabyashachi Mishra.
\newblock {Benchmarking of different optimizers in the variational quantum algorithms for applications in quantum chemistry}.
\newblock {\em The Journal of Chemical Physics}, 159(4):044117, 07 2023.

\bibitem{sack2024large}
Stefan~H Sack and Daniel~J Egger.
\newblock {Large-scale quantum approximate optimization on nonplanar graphs with machine learning noise mitigation}.
\newblock {\em Physical Review Research}, 6(1):013223, 2024.

\bibitem{ibmtime}
IBM.
\newblock Estimate job run time - ibm quantum documentation.

\end{thebibliography}

\end{document}